\documentclass[doublecol, figures]{epl2} 
\usepackage{amsmath, amsfonts, amssymb}
\usepackage[urlcolor=cyan]{hyperref}
\usepackage{cases}
\usepackage{mathtools}
\usepackage{soul}
\usepackage{upgreek} 
\usepackage{dsfont}
\usepackage{lmodern}
\usepackage{enumitem} 
\usepackage[dvipsnames]{xcolor}
\usepackage{soul}
\usepackage{tikz}
\newcommand{\nn}{\nonumber}
\newcommand{\be}{\begin{equation}}
\newcommand{\ee}{\end{equation}}
\newcommand{\bea}{\begin{eqnarray}}
\newcommand{\eea}{\end{eqnarray}}

\newcommand{\I}{\ensuremath{\mathbf{i}}}

\renewcommand{\geq}{\geqslant}

\title{Steady state of the KPZ equation on an interval and Liouville quantum mechanics}
\shorttitle{Steady state of the KPZ equation on an interval and Liouville quantum mechanics} %Insert here a short version of the title if it exceeds 70 characters

\author{Guillaume Barraquand\inst{1} \and Pierre Le Doussal\inst{1}}
\shortauthor{G. Barraquand and P. Le Doussal}

\institute{                    
  \inst{1} Laboratoire  de  Physique  de  l'\'Ecole  Normale  Sup\'erieure,  ENS, CNRS,  Universit\'e  PSL,   Sorbonne  Universit\'e,  Universit\'e de  Paris, 24 rue Lhomond, 75231 Paris, France.
}

\abstract{We obtain a simple formula for the stationary measure of the height field evolving according to the Kardar-Parisi-Zhang equation on the interval $[0,L]$ with general Neumann type boundary conditions and any interval size. This is achieved using the recent results of Corwin and Knizel (arXiv:2103.12253) together with Liouville quantum mechanics. Our formula allows to easily determine the stationary measure in various limits: KPZ fixed point on an interval, half-line KPZ equation, KPZ fixed point on a half-line, as well as the Edwards-Wilkinson equation on an interval.}

\begin{document}

\maketitle

\section{Introduction} The Kardar-Parisi-Zhang (KPZ) equation \cite{KPZ} describes the stochastic growth of a continuum interface driven by white noise.
In one dimension it is at the center of the so-called KPZ class which contains a number of well-studied models sharing the same universal behavior at large scale. For all these models one can define a height field.
For example, in particle transport models
such as the asymmetric simple exclusion process (ASEP) on a lattice, the local density is a discrete analog to the height gradient
\cite{derrida1998exactly, spohn2012large}. In the limit of weak asymmetry, ASEP converges \cite{bertini1997stochastic}, upon rescaling space and time
to the KPZ equation. In the large scale limit, all models in the KPZ class (in particular ASEP and the KPZ equation) are expected to converge to a universal process called the KPZ fixed point \cite{matetski2016kpz, corwin2015renormalization}. Note that the KPZ fixed point is universal with respect to the microscopic dynamics but still depends on the geometry of the space considered (full-line, half-line, circle, segment with boundary conditions).

An important question is the nature of the steady state. While the global height grows linearly in time with non trivial
$t^{1/3}$ fluctuations, the height gradient, or the height differences between any two points, will reach a stationary distribution. 
It was noticed long ago \cite{forster1977large,KPZ,parisi1990replica}
that the KPZ equation on the full line admits the Brownian motion as a stationary measure. It was proved rigorously in \cite{bertini1997stochastic, funaki2015kpz}, and in \cite{hairer2018strong} for periodic boundary conditions. 
For ASEP, stationary measures 
were studied on the full and half-line \cite{liggett1975ergodic, andjel1982invariant}  and 
exact formulas were obtained  on an interval using the matrix product ansatz \cite{derrida1993exact}. 
The large scale limit of the stationary measures for  ASEP on an interval was studied in \cite{derrida2004asymmetric,bryc2019limit}.  The processes obtained there as a limit can be described in terms of textbook stochastic processes such as  Brownian motions, excursions and meanders, and they should correspond to stationary measures of the KPZ fixed point on an interval.

 For the KPZ equation, while the stationary measures are simply Brownian in the full-line and circle case, the situation is more complicated (not translation invariant, not Gaussian) in the cases of the half-line and the interval. One typically imposes Neumann type boundary conditions (that is, we fix the derivative of the height field at the boundary) so that stationary measures depend on boundary parameters and involve more complicated stochastic processes (see below). 
For the KPZ equation on the half-line with Neumann type boundary condition, it can be shown \cite{barraquand2020half} that a Brownian motion with an appropriate drift is stationary (the drift must be proportional to the boundary parameter). 
This specific stationary measure was studied in \cite{barraquand2021kpz} for the equivalent directed polymer problem
for which the boundary parameter measures the attractiveness of the wall.  
However, based on the analysis of stationary measures of ASEP on a half-line \cite{liggett1975ergodic}, it was expected that more complicated stationary measures for the KPZ equation also exist.

The question of the stationary measure for the KPZ equation on the interval $[0,L]$ has also remained open. 
In a recent breakthrough, Corwin and Knizel obtained \cite{corwin2021stationary} an explicit formula for the Laplace transform of the stationary height distribution (for $L=1$, and for some range of parameters).  This Laplace transform formula relates the stationary measure to an auxiliary stochastic  process called continuous dual Hahn process. This construction corresponds to the KPZ equation limit\footnote{The limit from ASEP to KPZ was previously investigated in \cite{corwin2016open}. Remark 2.11 therein explains how to rescale ASEP boundary parameters to obtain the boundary conditions for the KPZ equation.} of formulas with a similar structure obtained in \cite{bryc2017asymmetric} of formulas with a similar structure obtained in \cite{bryc2017asymmetric} for the stationary measure of ASEP.  
However, obtaining a characterization of the process allowing to study properties of the stationary measure remains a challenge. This amounts to invert the complicated Laplace transforms in \cite{corwin2021stationary}.

In this Letter we obtain a simple formula for the stationary measure of the KPZ equation on an interval of any size $L$,
with general Neumann type boundary conditions. Our result is particularly convenient to study various limits as $L\to +\infty$. In particular we recover  the phase diagram for stationary measures of the KPZ fixed point on an interval, we obtain new crossover regimes near the critical point, and  we study the stationary measures of the KPZ equation on a half-line and their large scale limits. 
We unveil and exploit a surprising connection to Liouville quantum mechanics (LQM), i.e.
the Schrodinger equation in an exponential potential. The LQM, which is the 1D limit of the 2D Liouville field
theory, notably allows to study the statistics of exponential functionals of the Brownian motion,
see \cite{TexierComtetSUSY} for a short review.
As such,  it appears in several areas of physics such as diffusion in 1D random media
\cite{BouchaudPLD1990,monthus1994flux,broderix1995thermal,TexierComtetSUSY,comtet1998exponential,monthus2002localization,nagar2006strong},
multifractal eigenfunctions of random Schrodinger and Dirac operators 
\cite{kolokolov1994spatial,kolokolov1993method,TexierComtetSUSY,shelton1998effective,quinn2015scaling}, 
diffusion in the hyperbolic plane \cite{comtet1987landau,comtet1996diffusion,ikeda1999brownian},
and more recently, and strikingly, in quantum chaos and its relation to gravity \cite{bagrets2016sachdev,mertens2017solving}. 
LQM was recently used to obtain multipoint observables for the stationary KPZ equation in a half-space (see Supp. Mat. in \cite{barraquand2021kpz}). 

\bigskip 

\section{Model} The KPZ equation for the height field $h(x,t)$ reads
\be \label{eq:KPZ}
\partial_t h(x,t) = \nu \partial_x^2 h + \frac{\lambda}{2} (\partial_x h)^2 + \sqrt{D} \xi(x,t)
\ee
where $\xi(x,t)$ is a standard space-time white noise.
We use space-time units so that $\nu=1$ and $\lambda=D=2$
\footnote{The units chosen in \cite{corwin2021stationary} amount to rescaling our time as $t \to 2 t$,
immaterial in the steady state.}. Here we study the problem 
on the interval so that  \eqref{eq:KPZ} holds for $0<x<L$.
The solution is defined from the Cole-Hopf mapping $h(x,t)=\log Z(x,t)$,
where $Z(x,t)$ equals the partition sum of a continuum directed polymer with endpoint at $(x,t)$ in a random
potential $- \sqrt{2} \xi(x,t)$.
It satisfies the stochastic heat equation (SHE)
\be
\partial_t Z(x,t) = \partial^2_{x} Z(x,t) + \sqrt{2} Z(x,t) \xi(x,t), \, x \in [0,L]
\label{eq:mSHEhalf-space}
\ee 
in the Ito sense,
with Robin boundary conditions
\begin{equation}
\partial_x Z(x,t)|_{x=0}=A Z(0,t), \;\; 
\partial_x Z(x,t)|_{x=L}=- B Z(L,t),
\label{eq:boundarycondition}
\end{equation}
and it will be convenient to define boundary parameters $u=A+1/2$ and $v=B+1/2$. Although $Z(x,t)$, $t>0$, is not differentiable, the standard way to understand \eqref{eq:boundarycondition} is to impose these conditions on the heat kernel \cite{corwin2016open}, or through  a path integral as in 
\cite{borodin2016directed,deNardisPLDTT}. For the DP, $A>0$ corresponds to a repulsive wall and $A<0$ an attractive one, 
and similarly for $B$ at $x=L$.

\section{Main result} Our main result is the prediction that the KPZ random height profile in the stationary state, denoted $\left\lbrace  H(x) \right\rbrace_{x\in [0,L]}$, can be 
written as the sum of two independent random fields
\be \label{H} 
H(x) - H(0) = \frac{1}{\sqrt{2}} W(x) + X(x)
\ee 
where $W(x)$ is a one sided Brownian motion (i.e. with $W(0)=0$ and $W(L)$ free) 
and the probability distribution of the process $X(x)$ is given by the path integral measure
\be  \label{eq:path} 
\frac{\mathcal D X}{{\cal Z}_{u,v}} e^{ -  \int_0^L dx  (\frac{dX(x)}{dx})^2  }  e^{-2 v X(L)}\left( \int_0^L \mathrm dx\; e^{-2 X(x)} \right)^{-(u+v)}
\ee 
with $X(0)=0$ and $X(L)$ free, and ${\cal Z}_{u,v}$ a normalization such that ${\cal Z}_{0,0}=1$. For the choice $v=-u$,  $H(x) +vx$ is thus simply a standard Brownian motion. 
The Brownian motion with drift $u$ is also stationary for the half-space KPZ equation \cite{barraquand2020half}, and on the segment it arises only when the two boundary conditions are compatible. 

In mathematical terms, $X$ is a continuous stochastic process on $[0,L]$ whose measure $\mathbb P_X$ is absolutely continuous with respect to that of a Brownian motion with diffusion coefficient $1/2$, denoted $\mathbb P_B$, with Radon-Nikodym derivative  $\frac{d\mathbb P_X}{d\mathbb P_B}= \frac{1}{\mathcal Z_{u,v}}e^{-\mathcal E_{u,v}(X)}$ where 
%\begin{equation}e^{-\mathcal E (X)} = \left(  \int_0^L \mathrm dx e^{-2X(x)}\right)^{-u}  \left( \int_0^L \mathrm dx e^{2X(L)-2X(x)}\right)^{-v}\end{equation}
\begin{multline}\mathcal E_{u,v} (X) = \\  u \log  \left( \int_0^L \mathrm dx e^{-2X(x)}\right)+ v\log  \left(\int_0^L \mathrm dx e^{2X(L)-2X(x)}\right). 
\label{eq:RadonEnergy}
\end{multline}
This is a mere reformulation of  \eqref{eq:path} in a more symmetric form so that it becomes apparent that the process is left invariant after reversing space and exchanging $u,v$. This reformulation simply means that for any continuous and bounded functional $F$ of the process $X=\left\lbrace  X(x) \right\rbrace_{x\in [0,L]}$, 
\begin{equation}
    \mathbb E_X \left[ F\left(X\right) \right] = \frac{1}{\mathcal Z_{u,v}} \mathbb E_B \left[ F\left(B\right)e^{-\mathcal E_{u,v}(B)} \right],
    \label{eq:functionals}
\end{equation}
where in the R.H.S., $B=\left\lbrace  B(x) \right\rbrace_{x\in [0,L]}$ is a Brownian motion with diffusion coefficient $1/2$. 
Surprisingly the measure defined here in \eqref{eq:path} as the steady state of the KPZ equation has been already introduced and
studied in a work of Hariya and Yor in a different context \cite{hariya2004limiting}. We will use some of their results below. 

\section{Liouville quantum mechanics}
The Liouville Hamiltonian $\hat H$ on the real axis $U \in \mathbb{R}$ is defined as
\be \label{eq:HLiouville} 
\hat H = - \frac{1}{4} \frac{d^2}{dU^2} + e^{- 2 U} 
\ee 
Its eigenfunctions $|k \rangle$, in coordinate basis $\psi_k(U) = \langle U | k \rangle$,
can be chosen real and indexed by $k \geq 0$ with 
\be 
 \hat H \psi_k(U) = \frac{k^2}{4} \psi_k(U) ,
\quad  \quad \psi_k(U)= N_k K_{i k}( 2 e^{- U}).
\ee 
where  %$N_k =   \frac{1}{\pi} \sqrt{2 k \sinh( \pi k) }$ and 
$N_k^2= \frac{2}{\pi \Gamma(i k) \Gamma(- i k)}$ and $K_{\I k}$ is a modified Bessel function..
They form a continuum orthonormal basis with $\langle k | k' \rangle=\delta(k-k')$
with $\int_0^{+\infty} dk |k \rangle \langle k|=I$. 
%$k \sinh \pi k = \pi/(\Gamma(i k) \Gamma(- i k))$
We will need the matrix elements of the operator 
\be \label{defUhat} 
e^{- 2 \alpha \hat U } =\int_{\mathbb{R}} dU e^{- 2 \alpha U} |U \rangle \langle U|
\ee
which read
\bea \label{matrixelements} 
&& \langle k | e^{- 2 \alpha\hat U } | k' \rangle %= \frac{2}{\pi^2} \sqrt{k k' \sinh(\pi k) \sinh(\pi k') } 
= N_k N_{k'} \int_0^{+\infty} \frac{r^{2 \alpha} dr}{r} K_{ik}(2 r) K_{ik'}(2 r) \nn \\
&& 
%= \frac{1}{4 \pi \Gamma (2 \alpha )} \sqrt{ \frac{1}{\Gamma(i k) \Gamma(-i k) \Gamma(i k') \Gamma(-i k')} } 
= \frac{N_k N_{k'}}{8 \Gamma (2 \alpha )} \Gamma_4\left(\alpha \pm  \frac{i k}{2} \pm  \frac{i k'}{2}\right)
\eea
for any $\alpha>0$, where $\Gamma_4(\alpha \pm x \pm y) := \prod_{\sigma,\tau=\pm 1} \Gamma(a+\sigma x+\tau y)$
is a product of four Gamma functions.
\\
%\Gamma(\alpha + \frac{i k}{2} +  \frac{i k'}{2})\Gamma(\alpha - \frac{i k}{2} +  \frac{i k'}{2})
%\Gamma(\alpha + \frac{i k}{2} -  \frac{i k'}{2}) \Gamma(\alpha - \frac{i k}{2} -  \frac{i k'}{2}) \nn
%\eea

\section{Multipoint Laplace transform} 
We now compute, using the LQM, the multi-point Laplace transform (LT) of the distribution of $H(x)$, as defined in \eqref{H}, \eqref{eq:path}. For the moment, we restrict ourselves to $u,v>0$, and we will discuss the general case later. 
We consider the increasing sequence of points
\be 
x_0=0 < x_1 <\dots < x_m <x_{m+1}=L
\label{eq:defxi}
\ee
Since $W(x)$ is simply a Brownian independent of $X$ the following multipoint expectation, with parameters $\vec s=\lbrace 2u > s_1>\dots>s_m>s_{m+1}=0\rbrace$, takes the form
\be \label{eq:multilaplace} 
\mathbb{E}[ e^{- \sum_{j=1}^m s_j (H(x_j) - H(x_{j-1}))} ] =  e^{ \frac{1}{4} \sum_{j=1}^{m+1} s_j^2 (x_j-x_{j-1}) }
\frac{J(\vec s)}{J(0)}
\ee 
where $J(\vec s)$ is the following expectation over the process $X(x)$, from \eqref{eq:path}, with $X_0=0$ 
\begin{multline}\label{eq:Jstart} 
 J(\vec s)= \left[\prod_{j=1}^{m+1} \int_{\mathbb{R}} dX_j\right]  e^{- \sum_{j=1}^m s_j (X_j - X_{j-1}) -2 v X_{m+1}}    \\
\times \prod_{j=1}^{m+1} \int_{X(x_{j-1})=X_{j-1}}^{X(x_{j})=X_{j}}  \hspace{-.5cm}\mathcal D X e^{- \int_{x_{j-1}}^{x_j} dx (\frac{d X(x)}{dx})^2} 
Z_L[X]^{-(u+v)} 
\end{multline}
where $Z_L[X] := \int_0^L e^{-2 X(x)}$. Now we insert in the integrand the following representation 
\be 
Z_L[X]^{-(u+v)}= \frac{2}{\Gamma(u+v)} \int_{\mathbb{R}} dU_0 \, e^{-2 U_0(u+v)- e^{-2 U_0} Z_L[X]}
\ee
Performing the change of variable  $U_j=X_j+U_0$, and defining the process $U(x) = X(x) + U_0 $, the
RHS of \eqref{eq:Jstart}, i.e. $J(\vec s)$, takes the form
\begin{multline}  
 \frac{2}{\Gamma(u+v)} \prod_{j=0}^{m+1} \int_{\mathbb{R}} d U_j
 e^{- \sum_{j=1}^m s_j (U_j - U_{j-1}) -2 v U_{m+1} -2 u U_0} \nonumber 
 \\
\times 
\prod_{j=1}^{m+1} \int_{U(x_{j-1})=U_{j-1}}^{U(x_{j})=U_{j}} \mathcal D U e^{- \int_{x_{j-1}}^{x_j} dx [ (\frac{dU(x)}{dx})^2 +  e^{-2 U(x)}] }
\end{multline}
In the last line we recognize the path integral representation of the imaginary time Green's function of the Liouville Hamiltonian
$\hat H$ in \eqref{eq:HLiouville}. From the Feynman-Kac formula
\begin{multline} 
 \int_{U(a)=U}^{U(b)=U'} \mathcal D U e^{- \int_{a}^{b} dx \left[ (\frac{dU(x)}{dx})^2 +  e^{-2 U(x)}\right] } \\
 = \langle U' | e^{- (b-a) \hat H} | U \rangle \nn 
\end{multline} 
Hence we can rewrite $J(\vec s)=\frac{2}{\Gamma(u+v)} \tilde J(\vec s) $ with
\begin{multline} \label{a1} 
 \tilde J(\vec s) = 
\left[ \prod_{j=0}^{m+1} \int dU_j \right] e^{-2 v U_{m+1} - (2 u - s_1) U_0} \\
% \times 
\prod_{j=1}^m e^{-(s_{j}-s_{j+1}) U_j } 
\prod_{j=1}^{m+1} \langle U_{j} | e^{- (x_{j}-x_{j-1}) \hat H } |U_{j-1} \rangle.
\end{multline} 
Inserting the spectral decomposition
\be
e^{- (x_{j}-x_{j-1}) \hat H } = \int_0^{+\infty} dk_j e^{- \frac{1}{4} (x_{j}-x_{j-1}) k_j^2 } |k_j \rangle \langle k_j|
\ee 
and using the definition \eqref{defUhat},
the RHS of \eqref{a1} becomes
\begin{multline}   \label{RHS1} 
\!\! \! \! \int_{\mathbb{R}^2} \hspace{-.2cm} dU_0 dU_{m+1} \hspace{-.2cm}\prod_{j=1}^{m+1} \int_0^{+\infty} \hspace{-.5cm} dk_j
e^{-2 v U_{m+1} -  (2u-s_1) U_0} \langle U_{m+1} | k_{m+1} \rangle   \\
 \times  
\langle k_1 | U_0 \rangle
\prod_{j=1}^m \langle k_{j+1} | e^{-(s_j-s_{j+1}) \hat U } | k_{j} \rangle   e^{ - \sum_{j=1}^{m+1}  (x_j-x_{j-1}) \frac{k_j^2}{4} } 
\end{multline}
Now we use that for $w>0$
\bea \label{id1} 
\int_{\mathbb R} dU e^{- 2 w U} \langle U | k \rangle  &=& 
 N_k \int_{\mathbb R} dU e^{- 2 w U} K_{\I k}(2 e^{-U}) \nonumber \\ 
& =& \frac{N_k}{4} \left\vert \Gamma\left( w + \frac{\I k}{2}\right) \right\vert^2, 
\eea
and $\langle U | k \rangle=\langle k | U \rangle$ to integrate \eqref{RHS1}
w.r.t. $U_0$ and $U_{m+1}$. We can then substitute the matrix elements 
$\langle k_{j+1} | e^{-(s_j-s_{j+1}) \hat U } | k_{j} \rangle$ by their explicit
expressions from \eqref{matrixelements}. This leads to 
\begin{multline}  \label{eq:finalJ} 
\tilde J(\vec s) = \\  \frac{1}{2}  \prod_{j=1}^{m+1} \int_0^{+\infty} \hspace{-.5cm} \frac{dk_j}{4 \pi |\Gamma(i k_j)|^2} 
\prod_{j=1}^m \frac{\Gamma_4\left(\frac{s_j-s_{j+1}}{2} \pm \frac{i k_j}{2} \pm \frac{i k_{j+1}}{2}\right)}{\Gamma(s_j-s_{j+1}) } \\
 \times 
\left\vert \Gamma\left(u- \tfrac{s_1}{2} + \tfrac{i k_1}{2}\right) \Gamma\left(v + \tfrac{i k_{m+1}}{2}\right)\right\vert^2 \!\! e^{ \sum_{j=1}^{m+1}  \frac{-k_j^2}{4}(x_j-x_{j-1}) } .
\end{multline}

\section{Comparison with the result of Corwin-Knizel} 
Now, we explain why our formula for the stationary measure in \eqref{H} and \eqref{eq:path} is equivalent to the Laplace transform formula of Corwin-Knizel \cite{corwin2021stationary} for $L=1$.  Again, we assume that $u,v>0$ with $u<1$ (though \cite{corwin2021stationary} provides formulas for the whole range $u+v>0$). The Laplace transform of the stationary height $H(x)$ is expressed in \cite{corwin2021stationary}  in terms of a Markov process $\lbrace  \mathbb{T}_{s} \rbrace_{s \in [0,2u)}$ with values on $\mathbb{R}_+$, called the
continuous dual Hahn process (CDHP) 
%($C_{u,v}=\min(2 u,2)$ for $u>0$,$C_{u,v}=2$ for $u\leq 0$). 
The transition probability for $\mathbb{T}_{s_{j}}=t_{j}$
given $\mathbb{T}_{s_{j+1}}=t_{j+1}$, with $s_j>s_{j+1}$, is given by 
\begin{multline}
 p_{s_{j+1},s_j}(t_{j+1},t_j) = \frac{1}{8 \pi} \\
% = \frac{1}{8 \pi} 
 \times \frac{ |\Gamma(u- \frac{s_j}{2} + \frac{i}{2} \sqrt{t_j} )|^2 \Gamma_4(\frac{s_j-s_{j+1}}{2} \pm  \frac{i}{2} \sqrt{t_{j+1}} \pm \frac{i}{2} \sqrt{t_j} )  }
{ |\Gamma(u- \frac{s_{j+1}}{2} + \frac{i}{2} \sqrt{t_{j+1}} )|^2 \Gamma(s_j-s_{j+1}) \sqrt{t_j} |\Gamma(i \sqrt{t_j})|^2} \nonumber 
\end{multline}
and the marginal PDF of $\mathbb{T}_{s}=t$ at time $s$ is given by 
\begin{multline} 
p_s(t)= \\ \frac{(v+u)(v+u+1)}{8 \pi} \frac{ |\Gamma(v + \frac{s}{2} + \frac{i \sqrt{t}}{2} ) 
\Gamma(u - \frac{s}{2} + \frac{i \sqrt{t}}{2} ) |^2 }{\sqrt{t} |\Gamma( i \sqrt{t})|^2 }
\end{multline}
Then the Laplace transform is obtained, for $2u> s_1>s_2>\dots>s_{m+1}=0$, as
\be
\mathbb{E}\left[ e^{- \sum_{j=1}^m s_j (H(x_j) - H(x_{j-1})} \right] =  e^{ \frac{1}{4} \sum_{j=1}^{m+1} s_j^2 (x_j-x_{j-1}) }
\frac{I(\vec s)}{I_0}
\label{eq:LaplacetransformCK}
\ee 
with (upon some rewriting of formula (1.12) in \cite{corwin2021stationary}) 
\begin{multline} 
 I(\vec s) = \left[\prod_{j=1}^{m+1} \int_0^{+\infty} dt_j\right] p_0(t_{m+1}) \prod_{j=1}^m p_{s_{j+1},s_j}(t_{j+1},t_j)\\
 \times e^{ - \frac{1}{4} \sum_{j=1}^{m+1} t_j (x_j - x_{j-1} )}
\end{multline} 
and $I_0= I(0)=\int dt e^{- \frac{1}{4} t} p_0(t)$. It is now a simple exercise to check, using the change
of variables $t_j=k_j^2$,
that the above formula implies that 
\be 
I(\vec s)= 2 (u+v) (u+v+1) \tilde J(\vec s) = \Gamma(u+v+2) J(\vec s) 
\ee 
Since the prefactor cancels in the ratio, the multipoint Laplace transform \eqref{eq:multilaplace} of our formula \eqref{H}, \eqref{eq:path}  written more explicitly in \eqref{eq:finalJ}, 
coincides with the result of \cite{corwin2021stationary} in the domain $u,v>0$ and for $L=1$.

We may now extend the result of Corwin-Knizel in two directions. First, we may assume that $L$ is arbitrary, and reproduce the analysis of \cite{corwin2021stationary} with points $x_i$ as in \eqref{eq:defxi} using the same scalings as in \cite{corwin2021stationary} starting from the ASEP model on $NL$ sites and arriving at the same formula \eqref{eq:LaplacetransformCK}. This shows why \eqref{H},\eqref{eq:path} are correct for any $L>0$ and not only $L=1$ as considered in \cite{corwin2021stationary}. Then, we may extend the range of parameters $u,v$. We expect that the distribution of stationary measures depend analytically on $u,v$ for finite $L$.  However, performing a direct analytic continuation on \eqref{eq:LaplacetransformCK} is intricate and would involve many residues. This is why it is useful to rewrite, through LQM,  the results of Corwin-Knizel as in \eqref{H}, \eqref{eq:path}  which depends on the parameters in an analytic way  for all $u,v$.

Thus, from now on we will assume that our result for the stationary probability \eqref{eq:path}, \eqref{H} 
holds for any $u,v$, and explore the consequences.

Remark: Eq. \eqref{eq:LaplacetransformCK} relates the Laplace transform of the KPZ height field under the stationary measure to the Laplace transform of another Markov process, the CDHP, 
 where the role of time/space parameters and Laplace transform parameters are exchanged. 
The CDHP can be interpreted as living in the Fourier  space dual to the real $U$ space of LQM. 
A similar duality holds for Brownian excursions \cite{bryc2018dual} and can be obtained as a limit of \eqref{eq:LaplacetransformCK} as $L\to+\infty$,  as we shall see in the sequel. Note also that an analog of \eqref{eq:LaplacetransformCK} has been established for ASEP \cite{bryc2017asymmetric}, and it would be interesting to study the connection to discrete variants of LQM \cite{olshanetsky1994liouville}. It would be also very interesting to know if such dualities extend to other solvable models in the KPZ class.

\section{Limits and consequences} 

To study the various limits, we define the scaled processes $\widetilde H(\tilde x)$, $\widetilde W(\tilde x)$ and $\widetilde X(\tilde x)$, with $\tilde x=x/L \in [0,1]$ 
as
\be
H(x) = \sqrt{L} \widetilde H(\tilde x), \; W(x) = \sqrt{L} \widetilde W(\tilde x),\;  X(x) = \sqrt{L} \widetilde X(\tilde x)
\ee 
so that one has
\be 
\widetilde H(\tilde x) = \frac{1}{\sqrt{2}} \widetilde W(\tilde x) + \widetilde X(\tilde x)
\ee 
Clearly $\widetilde W(\tilde x)$ is also a standard one-sided Brownian motion. In order to impose Neumann type boundary conditions on $\widetilde H(\tilde x)$ (with slopes $\tilde u$, $-\tilde v$ respectively 
at each boundary) we scale boundary parameters as  
\begin{equation} u=\tilde u/\sqrt{L}, \;v=\tilde v/\sqrt{L}.
\end{equation}
The measure for $\widetilde X(\tilde x)$ can then
be written, up to a normalization, as $\mathcal D \widetilde X e^{ - S[\widetilde X] }$ with the action
\begin{multline}  \label{action0} 
 \! \! S[\widetilde X]= \int_0^1  d\tilde x  \left(\frac{d\widetilde X(\tilde x)}{d\tilde x}\right)^2 \!\!  + \frac{\tilde u}{\sqrt{L}} \log  \left(\int_0^1 \mathrm d\tilde x\; e^{-2 \sqrt{L} \widetilde X(\tilde x)}\right) \\
+ \frac{\tilde v}{\sqrt{L}} \log  \left(\int_0^1 \mathrm d\tilde x\; e^{2 \sqrt{L} (\widetilde X(1)-\widetilde X(\tilde x))}\right) 
\end{multline}
with $\widetilde X(0)=0$ and $\widetilde X(1)$ free. 

\subsection{Limit $L \to 0$} In this limit one recovers the Edwards-Wilkinson model. 
%To obtain a non trivial limit one scales $u=\tilde u/\sqrt{L}$ and $v=\tilde v/\sqrt{L}$ with $\tilde u=O(1)$ and $\tilde v=O(1)$.
 One sees that in \eqref{action0} we can expand up to
linear order in $\widetilde X$ in the logarithmic terms, and one finds that to leading order in $L$, i.e. to $L^0$, it becomes a Gaussian action with
a parabolic mean profile. One finds (see details in \cite{SM})
\be 
\widetilde H(\tilde x) - \widetilde H(0) \Longrightarrow \tilde u \tilde x - \frac{1}{2} (\tilde u+\tilde v) \tilde x^2 + \mathsf B(\tilde x)
\ee 
where $\mathsf B(\tilde x)$ is a standard Brownian motion. 

%\red{P: peut etre rajouter KPZ fixed point dans le titre de la section ci-dessous? insister un peu plus sur l'interet et 
%la nouveaute du resultat, on ne fait pas que retrouver Derrida !} 
\subsection{Limit $L \to +\infty$ (KPZ fixed point)}  Under the scalings considered above,  $\widetilde X$ converges to a probability measure proportional to
\be \label{measureFP} 
\mathcal D \widetilde X e^{ -  \int_0^1  d\tilde x  (\frac{d\tilde{X}(\tilde x)}{d\tilde x}+ \tilde v)^2  } 
e^{2  (\tilde u + \tilde v) \min_{\tilde x} \widetilde X(\tilde x)   } 
\ee 
with $\widetilde X(0)=0$. Hence it is a Brownian on $[0,1]$ with a non trivial Radon Nikodym derivative depending on $\tilde u, \tilde v$
(equivalently a Brownian with drift $-\tilde v$ with derivative depending only on $\tilde u+\tilde v$). The measure can also be rewritten in a more symmetric form as 
\be 
\mathcal D \widetilde X e^{ -  \int_0^1  d\tilde x  (\frac{d\tilde X(\tilde x)}{d\tilde x})^2  } 
e^{2  \tilde u  \min_{\tilde x} \lbrace \widetilde X(\tilde x) \rbrace  +2\tilde v \min_{\tilde x} \lbrace \widetilde X(\tilde x)-\widetilde X(1) \rbrace  }.  
\label{eq:crossoverweight}
\ee 
As detailed in \cite{SM} the measure \eqref{measureFP} can be studied 
using a limit of LQM, where the exponential potential is replaced by a hard wall. Accordingly all the above Laplace transform 
formula \eqref{eq:multilaplace}-\eqref{eq:finalJ}, are obtained for the rescaled process
and parameters by simply replacing $\Gamma(z) \to 1/z$.
The field $\widetilde H(\tilde x)$ should correspond to stationary measures of the  KPZ fixed point on the interval $[0,1]$ with boundary parameters $\tilde u, \tilde v$, and it is natural to predict that they arise as scaling limit of stationary measures of 
all models in the KPZ class on an interval. This is partially confirmed in some special cases that we study next, where we recover results obtained in  \cite{derrida2004asymmetric, bryc2019limit} for the large scale limit of ASEP stationary measures.

\section{Phase diagram} 
\begin{figure}
    \centering
    \begin{tikzpicture}[scale=1.4, every text node part/.style={align=center}]
    \draw[thick] (0,0) -- (2,2);
    \draw[thick, ->] (2,2) -- (2,4.1) node[anchor=north east] {$v$};
    \draw[thick, ->] (2,2) -- (4.1,2) node[anchor=north] {$u$};
    \draw[dashed, gray] (2,0) -- (2,2);
    \draw[dashed, gray] (0,2) -- (2,2);
    \draw[gray] (2,0.1) -- (2,-0.1) node[anchor=north] {$0$};
    \draw[gray] (0.1,2) -- (-0.1,2) node[anchor=east] {$0$};
    \draw (0.9,2.6) node{\footnotesize $\widetilde H(\tilde x) -u \sqrt{L}\tilde x$ \\ $\Downarrow$   \\ \footnotesize standard Brownian motion};
    \draw (2.5,1) node{\footnotesize $\widetilde H(\tilde x) +v \sqrt{L}\tilde x$ \\  $\Downarrow $  \\ \footnotesize standard Brownian motion}; 
    \draw (3,3) node{\footnotesize $\widetilde H(\tilde x) $ \\ $\Downarrow$\\   \footnotesize Brownian +  \\ \footnotesize Brownian excursion};  
    \draw (4.3,4.3) node{\footnotesize Brownian + \\ \footnotesize Brownian meander};    
    \draw[->] (3.3,4.2)  to[bend right] (2.05,3.5);    
    \draw[->] (4.3,3.9)  to[bend left] (3.5,2.05);    
    \end{tikzpicture}
    \caption{Phase diagram of the large-scale limit of stationary measures for the KPZ equation on the segment. On the three regions $u,v>0$ (maximal current phase), $v>u<0$ (low density phase) and $u>v<0$ (high density phase), we have indicated the nature of the stationary measure in the large scale limit.  For the directed polymer the phases are as 
    follows: For  $u,v>0$ the polymer is delocalized in the bulk, for $v<0,u>v$ it is bound to $x=L$, and for $u<0,u<v$ it is bound to $x=0$. 
    Exactly at the phase boundary $u=v$ 
    the polymer has probability $1/2$ to be bound to either side. } 
    \label{fig:phasediagram}
\end{figure}
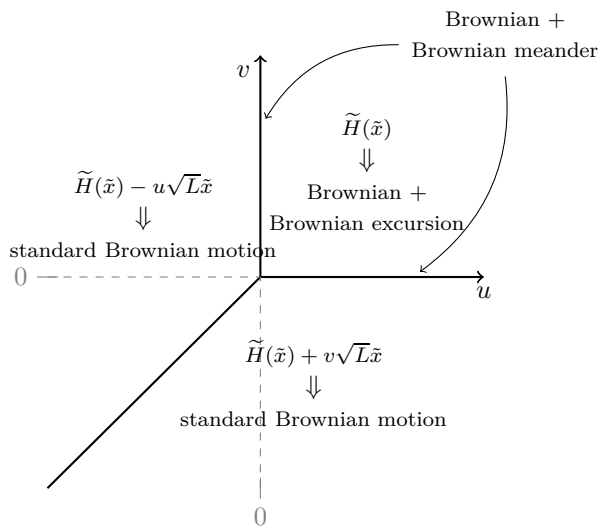

Now we study the phase diagram  in Fig. \ref{fig:phasediagram},  obtained in the $L\to\infty$ limit when $u,v$ are fixed (equivalently when $\tilde u, \tilde v$ go to $\pm\infty$).

\subsection{When $\tilde u, \tilde v \to +\infty$ ($u,v>0$)}

In that case the weight in \eqref{eq:crossoverweight} vanishes unless $\min_{\tilde x} \widetilde X(\tilde x)  = 0$ and $\min_{\tilde x} (\widetilde X(\tilde x) - \widetilde X(1)) =  0$, in which case the weight is $1$. 
The second identity  taken at $\tilde x=0$, together with the first taken at $\tilde x=1$ implies that $\widetilde X(1) =  0$. 
Thus, in the limit $\tilde u, \tilde v \to +\infty$, $\widetilde X(\tilde x) \Rightarrow \frac{1}{\sqrt{2}}  E(\tilde x)$, where $ E$
is a standard Brownian excursion (i.e. a Brownian bridge conditioned to stay positive), so that 
\be 
\widetilde H(\tilde x) \Longrightarrow \frac{1}{\sqrt{2}} \widetilde W(\tilde x) + \frac{1}{\sqrt{2}} E(\tilde x). 
\label{eq:MCphase}
\ee 
We recover the same process as in the large scale limit of TASEP \cite{derrida2004asymmetric} or ASEP \cite{bryc2019limit} invariant measures. 
This shall not be a surprise: TASEP, ASEP and the KPZ equation  all  converge at large time to the KPZ fixed point \cite{quastel2020convergence}, so that the process \eqref{eq:MCphase} describes the stationary measure of the KPZ fixed point in the so called maximal current phase for ASEP, that is with repulsive boundary conditions in terms of the directed polymer model. 

\subsection{When  $ \tilde u \to+\infty,  \tilde v=0$ or $ \tilde u=0,  \tilde v\to +\infty$ ($ u>0,  v=0$ or $ u=0,  v>0$)} 

When $ \tilde u \to+\infty,  \tilde v=0$ (equivalently $ u>0,  v=0$) the weight in \eqref{eq:crossoverweight} vanishes unless  $\min_{\tilde x} \widetilde X(\tilde x) = 0$, hence $\widetilde X$ is now a Brownian meander (Brownian motion conditioned to stay positive up to time $1$). We recover the same stationary process as in the large scale limit of ASEP stationary measures \cite{bryc2019limit}.
Similarly, when $ \tilde u=0,  \tilde v\to +\infty$ (equivalently $ u=0,  v>0$),  $\widetilde X(1)-\widetilde X(1-x)$ tends to a Brownian meander, and again, this matches with  \cite{bryc2019limit}.

\subsection{When $u$ or $v$ may be negative} By symmetry, we only need to consider the case where $v$ is negative. The measure for $X(x)$ in  \eqref{eq:path} 
is a Brownian measure with drift $-v$  weighted by  $Z_L[X]^{-(u+v)}$, 
%in its original form $e^{ - \int_0^L dx (\frac{dX}{dx} + v)^2 } Z_L(x)^{-(u+v)}$. 
%This is a Brownian with drift $-v$ weighted by $Z_L[X]^{-(u+v)}$,
where $Z_L[X]=\int_0^L dx e^{-2 X(x)}$. 
We may rewrite \eqref{eq:functionals}, up to a renormalisation,  as 
\begin{equation}
    \mathbb E_{B_{-v}} \left[ F(B_{-v}) Z_L[B_{-v}]^{-(u+v)}\right],
\end{equation}
where $B_{-v}$ denotes a Brownian with drift $-v$ and diffusion coefficient $1/2$. It is well-known that  $Z_L[B_{-v}]$ converges as $L\to\infty$ to
a finite random variable $Z_{\infty}[B_{-v}]$ distributed as an inverse Gamma law with shape parameter $-2v$ \cite{BouchaudPLD1990,Dufresne1990}, which   becomes asymptotically independent from the rescaled process $\widetilde X$. In order for $Z_{\infty}[B_{-v}]^{-(u+v)}$ to have a finite expectation, we need to assume that $-(u+v)<-2v$, that is $u>v$. Thus, we are considering the whole high density phase in Fig. \ref{fig:phasediagram}. At this point, we find that the resulting measure for $\widetilde H(\tilde x) + v\sqrt{L}\tilde x$ is simply the standard Brownian motion. In the low density phase where $u<0$ and $v>u$ (see Figure \ref{fig:phasediagram}), we deduce by symmetry that the limiting stationary process is a standard  Brownian motion with drift $u\sqrt{L}$.

\section{Half-line KPZ equation} 
Consider the KPZ equation \eqref{eq:KPZ} in $\mathbb R_+$, with boundary parameter $u$ at $x=0$. The stationary measures of the height field can be computed as the $L\to\infty$ limit of the stationary measures defined in \eqref{H} and \eqref{eq:path} and depending on parameters $u,v$. It turns out that this limit was studied in \cite{hariya2004limiting} (see also the review \cite{matsumoto2005exponential}). We will make considerable use of these results here. 
The set of stationary measures obtained in the $L\to\infty$ limit always depend on the boundary parameter $u$, and sometimes depend on the parameter $v$. 
When this is the case $v$ can be interpreted as minus the average drift of the process at infinity (see below).

In the low density phase ($u\leqslant 0$, $v\geqslant u$),  the stationary measure \eqref{H} simply converge as $L\to\infty$ to a standard Brownian motion with drift $u$ \cite{hariya2004limiting} (so that the limit does not depend on $v$). In the maximal current phase ($u\geqslant 0,v\geqslant 0$), the stationary measures converge \cite{hariya2004limiting} to a distribution that we denote  $\mathcal{HY}^{(0)}_{\gamma_u}$ -- the letters $\mathcal{HY}$ stand for Hariya-Yor. This is the distribution of the process  
\begin{equation}
H(x) = B^{(1)}(x)+ B^{(2)}(x)+\log\left( 1+\gamma_u \int_0^x e^{-2B^{(2)}(z)}dz\right),
\label{eq:halfspacemaximalcurrent}
\end{equation}
where $B^{(1)}(x), B^{(2)}(x)$ are two independent Brownian motions with diffusion coefficient $1/2$ and $\gamma_u$ is an independent Gamma random variable with shape parameter $u$.  Again, the limit does not depend on $v$. 
In the high-density phase ($v\leqslant 0, u\geqslant v$), the stationary measures converge \cite{hariya2004limiting} to a distribution that we denote  $\mathcal{HY}^{(-v)}_{\gamma_{u-v}}$. This is the distribution of the process  
\begin{multline}
H(x) = B^{(1)}(x)  + B^{(2)}(x) + vx \\ +\log\left( 1+\gamma_{u-v} \int_0^x e^{-2B^{(2)}(z)-2vz}dz\right),
\label{eq:halfspacehighdensity}
\end{multline}
where $B^{(1)}(x), B^{(2)}(x)$ are two independent Brownian motions with diffusion coefficient $1/2$ and $\gamma_{u-v}$ is an independent Gamma random variable with shape parameter $u-v$. In this phase, the limit depends on $v$, and we remark that the drift at infinity of the process \eqref{eq:halfspacehighdensity} is $-v$ (in the sense that $H(x)-H(0)\simeq -vx$ when $x\to+\infty$). 

%The KPZ equation in the half-space is obtained in the limit $L \to +\infty$ and looking at $H(x)$ at some fixed values $x=O(1)$. 
\begin{figure}
    \centering
    \begin{tikzpicture}[scale=1.3, every text node part/.style={align=center}]
    \draw[thick, black!20] (0,4) -- (4,0);
    \draw[thick] (0,0) -- (2,2);
    \draw[thick, ->] (2,2) -- (2,4.1) node[anchor=north east] {$v$ \\\footnotesize   drift \\ \footnotesize  parameter};
    \draw[thick, ->] (2,2) -- (4.1,2) node[anchor=north] {$u$ \\ \footnotesize  boundary parameter};
    \draw[dashed, gray] (2,0) -- (2,2);
    \draw[dashed, gray] (0,2) -- (2,2);
    \draw[gray] (2,0.1) -- (2,-0.1) node[anchor=north] {$0$};
    \draw[gray] (0.1,2) -- (-0.1,2) node[anchor=east] {$0$};
    \draw (0.8,2.5) node{\footnotesize Standard Brownian \\\footnotesize motion with drift $u$};
    \draw (2.4,0.7) node{$\mathcal{HY}^{(-v)}_{\gamma_{u-v}}$ \\ \footnotesize defined in \eqref{eq:halfspacehighdensity}};
    \draw (3,3) node{$\mathcal{HY}^{(0)}_{\gamma_u}$ \\ \footnotesize defined in \eqref{eq:halfspacemaximalcurrent}};
    \end{tikzpicture}
    \caption{Phase diagram of stationary measures for the KPZ equation in the half-space $\mathbb R_+$ with boundary parameter $u$. The diagram means that if the initial condition has drift $-v$ at infinity, the height field should converge at large time  under mild assumptions to the stationary measure indicated in one of the three regions of the $(u,v)$ plane. In particular, if $u\leqslant 0$ and the drift at infinity is $0$, which includes the flat initial data, the height field will converge to a Brownian motion with drift $u$, as predicted in \cite{barraquand2021kpz}. Along the antidiagonal line $u=-v$, the stationary measure is always a Brownian motion with drift $u$, see \cite{SM}.} 
    \label{fig:half-space}
\end{figure}

We expect that for a large class of initial conditions with drift at infinity equal to $-v$, the height field will converge at large time (modulo a global shift) to one of the stationary measures that we have just described,  according to the phase diagram in Figure \ref{fig:half-space}. This prediction is based on an analogous convergence result at large time for ASEP on a half-line proved in \cite{liggett1975ergodic}, though the stationary measures were explicitly described much later \cite{derrida1993exact, grosskinsky2004phase}. 

\subsection{Large-scale limit} 
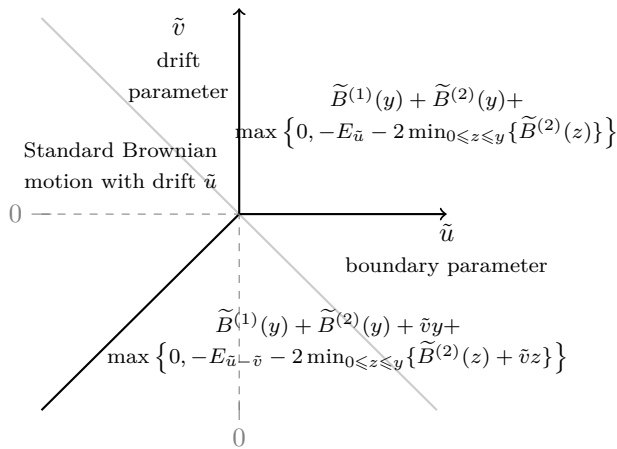
\begin{figure}
    \centering
    \begin{tikzpicture}[scale=1.3, every text node part/.style={align=center}]
    \draw[thick, black!20] (0,4) -- (4,0);
    \draw[thick] (0,0) -- (2,2);
    \draw[thick, ->] (2,2) -- (2,4.1) node[anchor=north east] {$\tilde v$ \\\footnotesize   drift \\\footnotesize  parameter};
    \draw[thick, ->] (2,2) -- (4.1,2) node[anchor=north] {$\tilde u$ \\ \footnotesize  boundary parameter};
    \draw[dashed, gray] (2,0) -- (2,2);
    \draw[dashed, gray] (0,2) -- (2,2);
    \draw[gray] (2,0.1) -- (2,-0.1) node[anchor=north] {$0$};
    \draw[gray] (0.1,2) -- (-0.1,2) node[anchor=east] {$0$};
    \draw (0.8,2.5) node{\footnotesize Standard  Brownian \\\footnotesize motion  with drift $\tilde u$};
    \draw (3,0.7) node{\footnotesize $\widetilde B^{(1)}(y)+ \widetilde B^{(2)}(y)+ \tilde v y + $ \\ \footnotesize $\max \left\lbrace 0, -E_{\tilde u-\tilde v} - 2 \min_{0\leqslant z\leqslant y}\lbrace \widetilde B^{(2)}(z)+\tilde v z\rbrace\right\rbrace $};
    \draw (3.9,3) node{\footnotesize $\widetilde B^{(1)}(y)+ \widetilde B^{(2)}(y)+$\\ \footnotesize $\max\left\lbrace 0, -E_{\tilde u}-2 \min_{0\leqslant z\leqslant y}\lbrace \widetilde B^{(2)}(z)\rbrace\right\rbrace $};
    \end{tikzpicture}
    \caption{Large scale limits of stationary measures of the half-space KPZ equation described in Fig. \ref{fig:half-space}. $\widetilde B^{(1)}(y)$ and $\widetilde B^{(2)}(y)$ denote independent Brownian motions with diffusion coefficient $1/2$. $E_{\tilde u}$ and $E_{\tilde u-\tilde v}$ denote independent exponential random variables with parameters $\tilde u$ and $\tilde u-\tilde v$. The parameters $\tilde u, \tilde v$ are rescalings of the parameters $u,v$ in Fig. \ref{fig:half-space} as explained in the Letter.}
    \label{fig:half-space2}
\end{figure}

Now that we have described the stationary measures of the KPZ equation on a half-line, it would be interesting to consider their large scale limit as $x$ goes to infinity. The processes obtained in this limit should be understood as stationary measures of the half-space KPZ fixed point, that is the universal process arising as scaling limit of all half-space models in the KPZ class. In particular, we conjecture that the large scale limit of half-line ASEP stationary measures \cite{grosskinsky2004phase} do converge to the same limit at large scale. Note that this half-space KPZ fixed point has not been defined rigorously (unlike the full-space situation), but its multipoint distributions for some initial conditions are known \cite{baik2018pfaffian, betea2020half}. The large scale limit of half-line KPZ equation stationary measures are described in the phase diagram of Fig. \ref{fig:half-space2}. 
We will let $x\to +\infty$ and define $\widetilde H(y) = \frac{1}{\sqrt{x}}H(xy)$ and scale the boundary parameter as $u=\frac{\tilde u}{\sqrt{x}}$. In the low-density phase, clearly, the Brownian motion with drift $u$ becomes a Brownian motion in the variable $y$  with drift $\tilde u$ in the scaling limit. In the maximal current phase ($\tilde u, \tilde v>0$), the large scale limit of \eqref{eq:halfspacemaximalcurrent} yields 
\begin{multline}
\widetilde H(y) = \widetilde B^{(1)}(y)+ \widetilde B^{(2)}(y) \\ 
+\max\left\lbrace 0, -E_{\tilde u}-2 \min_{0\leqslant z\leqslant y}\lbrace \widetilde B^{(2)}(z) \rbrace\right\rbrace ,
\label{eq:halfspacemaximalcurrentlimit}
\end{multline}
where $\widetilde B^{(1)}(y), \widetilde B^{(2)}(y)$ are two independent Brownian motions with diffusion coefficient $1/2$ and $E_{\tilde u}$ is an independent exponential random variable with rate parameter $\tilde u$. Here we have used that $\frac{-1}{\sqrt x}\log \gamma_{\tilde u/\sqrt{x}}$ converges to an exponential distribution with parameter $\tilde u$. In particular, when $\tilde u\to+\infty$, we obtain the sum of a Brownian motion and a Bessel 3 process \cite{pitman1975one} \footnote{We thank A. Comtet for an exchange on this point.}, that is a Brownian motion conditioned to remain positive on $[0,+\infty)$. 
In the high density phase ($\tilde v<0, \tilde u>\tilde v$), we scale $v=\frac{\tilde v}{\sqrt{x}}$ and  the large scale limit of \eqref{eq:halfspacehighdensity} yields 
\begin{multline}
\widetilde H(y) = \widetilde B^{(1)}(y)+ \widetilde B^{(2)}(y)+ \tilde v y \\ 
+\max\left\lbrace 0, -E_{\tilde u-\tilde v} - 2 \min_{0\leqslant z\leqslant y}\lbrace \widetilde B^{(2)}(z)+\tilde v z\rbrace\right\rbrace,
\label{eq:halfspacehighdensitylimit}
\end{multline}
where $\widetilde B^{(1)}(y), \widetilde B^{(2)}(y)$ are two independent Brownian motions with diffusion coefficient $1/2$ and $E_{\tilde u-\tilde v}$ is an independent exponential random variable with parameter $\tilde u-\tilde v$. Again, $-\tilde v$ represents the drift at infinity of the process $\widetilde H(y)$. When $\tilde u=-\tilde v$, the process \eqref{eq:halfspacehighdensitylimit} becomes a standard Brownian motion with drift $\tilde u$ as this was already the case before taking any limit, although this is not immediately obvious from \eqref{eq:halfspacehighdensitylimit}.

\section{Directed polymer endpoint} We obtain the endpoint probability for a 
very long polymer as ${\cal Q}(x)=e^{H(x)}/\int_0^L dx e^{H(x)}$,
where $H(x)$ is given in \eqref{H}. The statistics of the ratio ${\cal Q}(L)/{\cal Q}(0) = e^{\sqrt{\frac{L}{2}} G + Y}$, where
$G$ is a standard Gaussian random variable, 
requires only the PDF $P(Y)$ of $Y=X(L)-X(0)$, which in some special cases takes
a simple form \cite{SM}. For $u+v=-1$ one obtains
\be
P(Y) = \frac{e^{-(1+ 2 v) Y + \frac{L}{4}}}{2 {\cal Z}_{u,v}}  
   \left[\text{erf}\left(\frac{L-2 Y}{2 \sqrt{L}}\right)+\text{erf}\left(\frac{L+2 Y}{2
   \sqrt{L}}\right)\right]
\ee
with ${\cal Z}_{u,v} = \frac{e^{v^2 L}}{1+ 2 v} (e^{(1+2 v) L}-1)$.
At the transition point, $u=v=-1/2$, $P(Y)$ becomes uniform in
$[-L/2,L/2]$ consistent with the polymer being localized near either
boundary with probability $1/2$ (see \cite{SM} for details, and \cite{krug1994disorder} for an earlier work based on ASEP). 

\subsection{Note} While this work was near completion, the preprint \cite{BrycNew} appeared. This paper also performs a Laplace transform inversion of the result of \cite{corwin2021stationary}, although \cite{BrycNew} describes the process $X$ differently, as a Markov process with explicit transition probabilities. Both descriptions are equivalent: indeed, their formula \cite[(1.7)]{BrycNew} can also be read from \eqref{a1} above.

\smallskip 
\acknowledgements PLD acknowledges support from ANR grant ANR-17-CE30-0027-01 RaMaTraF. 

\bibliographystyle{eplbib_withtitles}
\bibliography{biblio1.bib}

\end{document}

\maketitle
\makeatletter
\newcounter {section}
\newcounter {subsection}[section]
\renewcommand \thesection {\@arabic \c@section .}
\renewcommand\thesubsection   {\thesection\@arabic\c@subsection}
\renewcommand\section{\@startsection {section}{1}{\z@}%
                                   {-3.5ex \@plus -1ex \@minus -.2ex}%
                                   {2.3ex \@plus.2ex}%
                                   {\normalfont\large\bfseries}}
\renewcommand\subsection{\@startsection{subsection}{2}{\z@}%
                                     {-3.25ex\@plus -1ex \@minus -.2ex}%
                                     {0.1ex}%
                                     {\normalfont\bfseries}}
\renewcommand\tableofcontents{%
%    \section*{\contentsname
%        \@mkboth{%
%           \MakeUppercase\contentsname}{\MakeUppercase\contentsname}}%
    \@starttoc{toc}%
    }
\renewenvironment{thebibliography}[1]{%
  \@startsection{section}{1}{0pt}{\epl@prebiblio}{\epl@postbiblio}%
  {}{\noindent\bf  References}%
      \def\and{\unskip\global\epl@gotandtrue{\normalfont\ and\ }\ignorespaces}%
      \list{\@biblabel{\@arabic\c@enumiv}}%
           {\settowidth\labelwidth{\@biblabel{#1}}%
            \leftmargin\labelwidth
            \advance\leftmargin\labelsep
            \@openbib@code
            \usecounter{enumiv}%
            \let\p@enumiv\@empty
            \renewcommand\theenumiv{\@arabic\c@enumiv}%
	    \parsep0pt
            \itemsep0pt
	    \small
      }%
      \sloppy
      \clubpenalty4000
      \@clubpenalty \clubpenalty
      \widowpenalty4000%
      \sfcode`\.\@m}
     {\def\@noitemerr
       {\@latex@warning{Empty `thebibliography' environment}}%
      \endlist}
\setcounter{tocdepth}{1}
\makeatother

\setcounter{secnumdepth}{2}
\tableofcontents

\vspace{1cm}

\section{Edwards-Wilkinson limit}
Consider the measure $\mathcal D \widetilde X e^{ - S[\widetilde X] }$ for the rescaled process, up to
a normalization, given in Eq.(28) in the Letter, and let us rewrite it in the equivalent form
\be 
S[\widetilde X]= \int_0^1  d\tilde x  \left(\frac{d\widetilde X(\tilde x)}{d\tilde x}\right)^2  + \frac{(\tilde u+ \tilde v)}{\sqrt{L}}  \log  \left( \sqrt{L}  \int_0^1 \mathrm d\tilde x\; e^{-2 \sqrt{L} \widetilde X(\tilde x)}\right)
+ 2 \tilde v \widetilde X(1),
\ee 
with $\widetilde X(0)=0$ and $\widetilde X(1)$ free. For $L\to 0$, $S[\widetilde X]$ can be expanded as 
\be 
S[\widetilde X]= \int_0^1  d\tilde x  \left(\frac{d\widetilde X(\tilde x)}{d\tilde x}\right)^2  + \frac{(\tilde u+ \tilde v)}{\sqrt{L}}  \log  \left( 1 - 2 \sqrt{L}  \int_0^1 \mathrm d\tilde x\; \widetilde X(\tilde x) 
+ 2 L \int_0^1 \mathrm d\tilde x \widetilde X(\tilde x)^2 + \dots \right) 
+ 2 \tilde v \widetilde X(1) , 
\ee 
where we do not keep track of constant terms which are absorbed in the normalization. To the leading order as $L \to 0$,  we obtain
\be 
S[\widetilde X]\simeq \int_0^1  d\tilde x  \left(\frac{d\widetilde X(\tilde x)}{d\tilde x}\right)^2  - 2 (\tilde u+ \tilde v) \int_0^1 \mathrm d\tilde x\; \widetilde X(\tilde x) 
+ 2 \tilde v \widetilde X(1) 
\ee 
which we can rewrite, using integration by parts (and up to an irrelevant constant) 
\be 
S[\widetilde X]\simeq \int_0^1  d\tilde x  \left(\frac{d\widetilde X(\tilde x)}{d\tilde x} - \tilde u + (\tilde u+\tilde v) \tilde x\right)^2  
\ee 
This is a Gaussian measure and one can write the convergence in distribution 
\be 
\widetilde X(x) \Longrightarrow \tilde u \tilde x - \frac{1}{2} (\tilde u+\tilde v) \tilde x^2 + \frac{1}{\sqrt{2}} \tilde {\sf B}(\tilde x) 
\ee 
where $\tilde {\sf B}(\tilde x)$ is a standard one sided Brownian motion with $\tilde {\sf B}(0)=0$. The mean profile of $\widetilde X(\tilde x)$ is thus quadratic 
with a curvature proportional to $-(\tilde u+\tilde v)$. Thus
the (rescaled) stationary height field $\widetilde H$ converges to a parabola plus a Brownian motion,  
\be 
 \widetilde H(\tilde x) \Longrightarrow \tilde u \tilde x - \frac{1}{2} (\tilde u+\tilde v) \tilde x^2 + {\sf B}(\tilde x)
\label{eq:EWstationary}
\ee 
where ${\sf B}(\tilde x)$ is another standard Brownian motion, as stated in the Letter. 

\begin{figure}
    \centering
    \begin{tikzpicture}[scale=8]
    \draw[thick, ->] (0,0) -- (1.1,0) node[anchor=south] {$\tilde x$};
    \draw[thick, -> ] (0,0) -- (0,0.25) node[anchor=west] {$\widetilde X(\tilde x)$};
    \draw[thick] (1,0.01) -- (1,-0.01) node[anchor=north] {$1$};
     \draw[thick] (0.,0.) -- (0.01,0.00766121) -- (0.02,0.0247122) -- (0.03,0.0367758) -- (0.04,0.0652824) -- (0.05,0.0693033) -- (0.06,0.0870228) -- (0.07,0.0922206) -- (0.08,0.101857) -- (0.09,0.109814) -- (0.1,0.119923) -- (0.11,0.123321) -- (0.12,0.143494) -- (0.13,0.133157) -- (0.14,0.141928) -- (0.15,0.146891) -- (0.16,0.151427) -- (0.17,0.174537) -- (0.18,0.184369) -- (0.19,0.19134) -- (0.2,0.186444) -- (0.21,0.181245) -- (0.22,0.178103) -- (0.23,0.193954) -- (0.24,0.191595) -- (0.25,0.193015) -- (0.26,0.198418) -- (0.27,0.201498) -- (0.28,0.213431) -- (0.29,0.21688) -- (0.3,0.229649) -- (0.31,0.233232) -- (0.32,0.227295) -- (0.33,0.223174) -- (0.34,0.230499) -- (0.35,0.226568) -- (0.36,0.234141) -- (0.37,0.24037) -- (0.38,0.244217) -- (0.39,0.225511) -- (0.4,0.229861) -- (0.41,0.224337) -- (0.42,0.237045) -- (0.43,0.243989) -- (0.44,0.221446) -- (0.45,0.218612) -- (0.46,0.20185) -- (0.47,0.22572) -- (0.48,0.234963) -- (0.49,0.238906) -- (0.5,0.246944) -- (0.51,0.244636) -- (0.52,0.266717) -- (0.53,0.268177) -- (0.54,0.256042) -- (0.55,0.253225) -- (0.56,0.257389) -- (0.57,0.256798) -- (0.58,0.261724) -- (0.59,0.253656) -- (0.6,0.249911) -- (0.61,0.256794) -- (0.62,0.269402) -- (0.63,0.268108) -- (0.64,0.2824) -- (0.65,0.289611) -- (0.66,0.262407) -- (0.67,0.258425) -- (0.68,0.268522) -- (0.69,0.270976) -- (0.7,0.253744) -- (0.71,0.252086) -- (0.72,0.270031) -- (0.73,0.262238) -- (0.74,0.247587) -- (0.75,0.252058) -- (0.76,0.241424) -- (0.77,0.246588) -- (0.78,0.229387) -- (0.79,0.224094) -- (0.8,0.218097) -- (0.81,0.230654) -- (0.82,0.23856) -- (0.83,0.224617) -- (0.84,0.239519) -- (0.85,0.212758) -- (0.86,0.202646) -- (0.87,0.201512) -- (0.88,0.204779) -- (0.89,0.197407) -- (0.9,0.190265) -- (0.91,0.177569) -- (0.92,0.169376) -- (0.93,0.137019) -- (0.94,0.128534) -- (0.95,0.128521) -- (0.96,0.0973639) -- (0.97,0.0881485) -- (0.98,0.0899677) -- (0.99,0.0831956) -- (1.,0.0818669); 
    \draw[gray, dashed, thick] (0.,0.) -- (0.01,0.0099) -- (0.02,0.0196) -- (0.03,0.0291) -- (0.04,0.0384) -- (0.05,0.0475) -- (0.06,0.0564) -- (0.07,0.0651) -- (0.08,0.0736) -- (0.09,0.0819) -- (0.1,0.09) -- (0.11,0.0979) -- (0.12,0.1056) -- (0.13,0.1131) -- (0.14,0.1204) -- (0.15,0.1275) -- (0.16,0.1344) -- (0.17,0.1411) -- (0.18,0.1476) -- (0.19,0.1539) -- (0.2,0.16) -- (0.21,0.1659) -- (0.22,0.1716) -- (0.23,0.1771) -- (0.24,0.1824) -- (0.25,0.1875) -- (0.26,0.1924) -- (0.27,0.1971) -- (0.28,0.2016) -- (0.29,0.2059) -- (0.3,0.21) -- (0.31,0.2139) -- (0.32,0.2176) -- (0.33,0.2211) -- (0.34,0.2244) -- (0.35,0.2275) -- (0.36,0.2304) -- (0.37,0.2331) -- (0.38,0.2356) -- (0.39,0.2379) -- (0.4,0.24) -- (0.41,0.2419) -- (0.42,0.2436) -- (0.43,0.2451) -- (0.44,0.2464) -- (0.45,0.2475) -- (0.46,0.2484) -- (0.47,0.2491) -- (0.48,0.2496) -- (0.49,0.2499) -- (0.5,0.25) -- (0.51,0.2499) -- (0.52,0.2496) -- (0.53,0.2491) -- (0.54,0.2484) -- (0.55,0.2475) -- (0.56,0.2464) -- (0.57,0.2451) -- (0.58,0.2436) -- (0.59,0.2419) -- (0.6,0.24) -- (0.61,0.2379) -- (0.62,0.2356) -- (0.63,0.2331) -- (0.64,0.2304) -- (0.65,0.2275) -- (0.66,0.2244) -- (0.67,0.2211) -- (0.68,0.2176) -- (0.69,0.2139) -- (0.7,0.21) -- (0.71,0.2059) -- (0.72,0.2016) -- (0.73,0.1971) -- (0.74,0.1924) -- (0.75,0.1875) -- (0.76,0.1824) -- (0.77,0.1771) -- (0.78,0.1716) -- (0.79,0.1659) -- (0.8,0.16) -- (0.81,0.1539) -- (0.82,0.1476) -- (0.83,0.1411) -- (0.84,0.1344) -- (0.85,0.1275) -- (0.86,0.1204) -- (0.87,0.1131) -- (0.88,0.1056) -- (0.89,0.0979) -- (0.9,0.09) -- (0.91,0.0819) -- (0.92,0.0736) -- (0.93,0.0651) -- (0.94,0.0564) -- (0.95,0.0475) -- (0.96,0.0384) -- (0.97,0.0291) -- (0.98,0.0196) -- (0.99,0.0099) -- (1.,0.);
    \end{tikzpicture}
    \caption{ Steady-state of the Edwards-Wilkinson equation on the interval $[0,1]$, depicted here for $\tilde u=\tilde v>0$. It is given by a parabola plus a standard Brownian motion -- see \eqref{eq:EWstationary}.} 
    \label{fig:my_label}
\end{figure}
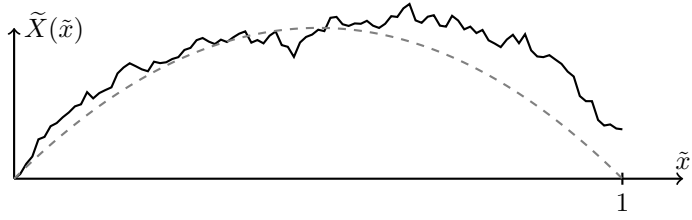

\medskip 
\textbf{Remark: } The fact that the RHS of \eqref{eq:EWstationary} is stationary for the EW equation can be shown by a direct calculation \cite{unpublished} : First, by substracting a parabola, one may reduce to the case $u=v=0$. Then, we observe that starting from Brownian initial data, the height is a Gaussian process. Finally, using the explicit form of the Green's function of the heat equation with Neumann boundary conditions, the spatial covariance  can be  computed explicitly and shown to be that of a Brownian motion. 

\section{Observables of the process $X$} 

Let us define, for any process $X(x)$ with $0\leq x\leq L$,
\be 
Z_L[X] = \int_0^L dx e^{-2 X(x)}.
\ee 
By definition,  the average of any observable $O[X]$ over the process $X(x)$,  defined in Eq. (5) of the Letter, 
can be written in the following path integral forms
\bea \label{average1}
\mathbb{E} \left[ O[X] \right] &=& \frac{1}{{\cal Z}_{u,v,L}} \int_{X(0)=0} \mathcal D X O[X] e^{ -  \int_0^L dx  (\frac{dX(x)}{dx})^2  }  e^{-2 v X(L)}  Z_L[X]^{-(u+v)}, \\
&=& \frac{e^{v^2 L}}{{\cal Z}_{u,v,L}} \int_{X(0)=0} \mathcal D X O[X] e^{ -  \int_0^L dx  (\frac{dX(x)}{dx} + v)^2  } Z_L[X]^{-(u+v)}.
\eea 
where ${\cal Z}_{u,v,L}$ is the normalisation. Equivalently, as in Eq. (7) of the Letter, it can be written as averages over the Brownian motion in two equivalent forms
\be \label{average2} 
 \mathbb{E} \left[ O[X] \right] = \frac{1}{{\cal Z}_{u,v,L}} \mathbb{E}_{B}  \left[ O[B]  e^{-2 v B(L)} Z_L[B]^{-(u+v)} \right]   = \frac{e^{v^2 L}}{{\cal Z}_{u,v,L}} \mathbb{E}_{B_{-v}} \left[ O[B_{-v}]  Z_L[B_{-v}]^{-(u+v)} \right], 
\ee 
where $\mathbb{E}_{B}$ denotes the average over the one-sided Brownian motion of diffusion constant $1/2$ (with $B(0)=0$), i.e. $B(x)= \frac{1}{\sqrt{2}} {\sf B}(x)$ 
where ${\sf B}(x)$ is the standard one-sided Brownian motion, and $\mathbb{E}_{B_{-v}}$ the average over the one-sided Brownian motion of diffusion constant $1/2$
with a drift $-v$, i.e. $B_{-v}(x)=\frac{1}{\sqrt{2}} {\mathsf B}(x)-v x$. 
The normalization is thus equal to the following expectations. 
\be \label{norm1} 
{\cal Z}_{u,v,L} = \mathbb{E}_{B}  \left[ e^{-2 v B(L)} Z_L[B]^{-(u+v)}  \right] 
= e^{v^2 L} \mathbb{E}_{B_{-v}} \left[ Z_L[B_{-v}]^{-(u+v)}  \right] 
\ee 
It admits a simple expression in some cases, we will give examples below.

\medskip 
It is interesting to note that the moments of $Z_L[X]$ for the process $X$ can be expressed from the normalization as (here $k$ can be any real number) 
\be
\mathbb{E} \left[ Z_L[X]^{k}  \right] = \frac{ {\cal Z}_{u-k,v} }{{\cal Z}_{u,v}}.
\ee 
A related property is that if we denote $Q_L^B(Z)$ the PDF of the variable $Z=Z_L[B_{-v}]$ for the Brownian $B_{-v}$ with diffusion coefficient $1/2$ and
drift $-v$, then the PDF $Q_L(Z)$ of $Z=Z_L[X]$ for the process $X(x)$ is simply
\be 
Q_L(Z) = C_{u,v,L} Q^B_L(Z) Z^{- (u+v)} , 
\ee 
where $C_{u,v,L}$ is a normalization. One simple application, discussed in the Letter, is for the case $v<0$. Then we know that $Z=Z_L[B_{-v}]$ 
has a limit distribution as $L \to +\infty$ given by an inverse Gamma law $1/\Gamma(-2 v,1)$
\be 
Q^B_{+\infty}(Z) = \frac{1}{\Gamma(-2 v)} \frac{1}{Z^{1-2 v}} e^{-1/Z} .
\ee 
If $u>v$, we see that $Z=Z_L[X]$ will also have a limit distribution for $L \to +\infty$ given by another inverse Gamma law $1/\Gamma(u- v,1)$
\be 
Q_{+\infty}(Z) = \frac{1}{\Gamma(u-v)} \frac{1}{Z^{1+u-v}} e^{-1/Z} .
\ee 
\\

{\bf One-point distribution of total height difference}. Consider here the height difference between the two boundaries of the interval. From Eq. (4) in the Letter  
we have
\bea 
H(L) - H(0) = \frac{\sqrt{L}}{\sqrt{2}} G + X(L)
\eea 
where $G$ is an independent unit Gaussian random variable. From \eqref{average2} 
choosing $O[X]=\delta(X(L)-Y)$, the PDF $P_L(Y)$ of the random variable $Y=X(L)$ is given by
\bea \label{onepoint} 
 P_L(Y) &=& \mathbb{E} \left[ \delta(X(L)-Y) \right]= \frac{1}{{\cal Z}_{u,v,L}} \mathbb{E}_{B}  \left[ \delta(B(L)-Y)  e^{-2 v B(L)} Z_L[B]^{-(u+v)} \right] \\
 &=& \frac{1}{{\cal Z}_{u,v,L}} e^{-2 v Y} \frac{e^{ - \frac{Y^2}{L} }}{\sqrt{\pi L}} \, \mathbb{E}_{B} \left[ Z_L[B]^{-(u+v)} \Big\vert B(L)=Y\right] \label{54} 
\eea 
where the expectation value is over a Brownian motion $B(x)$ with diffusion coefficient $1/2$ and  $B(0)=0$, conditioned to $B(L)=Y$.
We give some simple explicit examples below. 

An interesting property, which is a consequence of the form of the measure \eqref{average2}, 
is that the cumulants (respectively the moments) of the random variable $X(L)$ (i.e. the cumulants/moments of $P_L(Y)$ denoted $\langle Y^n \rangle_L^c$ and $\langle Y^n \rangle_L$ respectively) can be obtained from the normalisation.
Indeed setting $u+v=w$ fixed and taking successive derivatives w.r.t. $v$ one obtains
\bea \label{cumgenY} 
 \langle Y^n \rangle_L^c &=& \left.\left[ \left(-\tfrac{1}{2} \partial_v\right)^n \log {\cal Z}_{w-v,v,L} \right]\right\vert_{w=u+v}, \\
\langle Y^n \rangle_L &=& \left.\left[ \frac{1}{{\cal Z}_{w-v,v,L}}  \left(-\tfrac{1}{2} \partial_v\right)^n {\cal Z}_{w-v,v,L} \right]\right\vert_{w=u+v} .
\eea

\section{Some explicit formulas} 

In this section we use our main result, i.e. Eqs. (4) and (5) in the Letter,  to obtain some simple explicit formulas in special cases. 

\subsection{The Brownian case }
The simplest case is $u+v=0$, in which case one finds ${\cal Z}_{u,v}=e^{v^2 L}$ and
\bea
P_L(Y) = \frac{1}{\sqrt{\pi L}} e^{ - \frac{(Y+v L)^2}{L} } 
\eea 
which leads to $H(L)-H(0)= \frac{\sqrt{L}}{\sqrt{2}} G + \frac{\sqrt{L}}{\sqrt{2}} G' + u L = \sqrt{L} G'' + u L$, as expected
(here $G$ and $G'$ are independent unit Gaussian variables, and $G''$ another unit Gaussian variable). More generally
the process $X(x)$ has the same law as $B_{-v}(x)=B_u(x)$. 

\subsection{Normalisation}
There is one case where ${\cal Z}_{u,v,L}$ is trivial. For $u=1/2$ and $v=0$ one has from  \cite[Prop. 5.7]{matsumoto2005exponential}
\be
{\cal Z}_{u,v,L} = \mathbb{E}_B\left[ Z_L[B]^{-1/2} \right] = L^{-1/2}.
\ee 
For $u+v=-n$, where $n$ is a positive integer, it is possible to obtain explicit expressions. One has from \cite{monthus1994flux} (with $\mu=-2 v$, $\beta=2$, $\sigma=1/2$, $-\beta F_0=2 v$ )
or from  \cite[Thm 5.2]{matsumoto2005exponential} (with $\mu=2 v$),
\be  \label{normN} 
{\cal Z}_{u,v,L} = e^{v^2 L} \sum_{k=0}^n e^{L k (k + 2 v)} (-1)^{n-k} \binom{n}{k} (2 k + 2 v) \frac{\Gamma(k+2 v)}{\Gamma(n+1+k+2 v)}.
\ee 
For instance, for $n=1$, we have by \eqref{norm1} 
\be  \label{Norm1} 
{\cal Z}_{u,v,L} = \mathbb{E}_{B} \left[  \int_0^L dx e^{- 2 v B(L) -2 B(x)}  \right] = \int_0^L dx e^{(1+2 v) x + v^2 L} = \frac{e^{v^2 L}}{1+ 2 v} (e^{(1+2 v) L}-1)
\ee 
where the average is over a Brownian motion $B(x)$ with $B(0)=0$ and diffusion coefficient $1/2$.
 It is consistent with \eqref{normN} for $n=1$.

\subsection{PDF of the total height difference}
Similarly a direct calculation of $P_L(Y)$ is possible for $u+v=-n$, using \eqref{54}. 
To this aim let us define a Brownian bridge $B(x)$ such that $B(L)=Y$,  in terms of the standard Brownian bridge $\tilde {\sf B}(x)$,  as
\be 
B(x) = \frac{Y}{L} x + \frac{1}{\sqrt{2}} \tilde {\sf  B}(x) , \quad \tilde {\sf B}(x) = W(x) - \frac{x}{L} W(L) , \quad \mathbb{E} [ \tilde {\sf B}(x_1) \tilde {\sf B}(x_2) ] =
\min(x_1,x_2)- \frac{x_1 x_2}{L},
\ee 
where $W(x)$ is a standard Brownian with $W(0)=0$. 
\\

\textbf{Case $u+v=-1$.} One has
\be 
\mathbb{E}_{B} \left[\int_0^L dx e^{-2 B(x)}  \right]
= \int_0^L dx e^{- 2 \frac{Y}{L} x + x(1 - \frac{x}{L}) } = \frac{1}{2} \sqrt{\pi L} e^{\frac{(L-2 Y)^2}{4 L}}
   \left(\text{erf}\left(\frac{L-2 Y}{2 \sqrt{L}}\right)+\text{erf}\left(\frac{L+2 Y}{2
   \sqrt{L}}\right)\right)
\ee 
Inserting in \eqref{54} and using \eqref{Norm1} we obtain
\be  \label{PX} 
P_L(Y) = \frac{1}{2 {\cal Z}_{u,v}} e^{-(1+ 2 v) Y + \frac{L}{4}} 
   \left(\text{erf}\left(\frac{L-2 Y}{2 \sqrt{L}}\right)+\text{erf}\left(\frac{L+2 Y}{2
   \sqrt{L}}\right)\right).
\ee 
One can check that \eqref{PX} with \eqref{Norm1}
is normalized to unity $\int_{\mathbb{R}} dY P_L(Y)=1$. 
\bigskip 

On the transition line at the point $u=v=-1/2$,  we obtain a symmetric distribution for $Y= X(L)$
\be 
P_L(Y) = \frac{\text{erf}\left(\frac{L-2 Y}{2
   \sqrt{L}}\right)+\text{erf}\left(\frac{L+2 Y}{2 \sqrt{L}}\right)}{2 L}.
\ee 
It is easy to see that in the large $L$ limit, $P_L(Y)$ converges to the uniform distribution on the interval $[-L/2,L/2]$ with a boundary layer of
size $\sqrt{L}$ near the edges. As mentioned in the Letter, in the context of the directed polymer, this is consistent with the prediction
that the polymer is localized at either boundary with probability $1/2$. A possible scenario which would explain this limit distribution, is that the process $X(x)$ attains its minimum
at a random location $x^*$ uniformly distributed between $x=0$ and $x=L$, and the process is approximately (as $L$ goes to infinity) a straight line with slope $u$ on $[0, x^*]$ and a straight line 
with slope $-v$ on $[x^*,1]$. It would be interesting to confirm this scenario by computing multipoint distributions (see below). 

\textbf{Remark:} For TASEP, described as a zero temperature polymer model, a similar scenario was discovered in \cite{krug1994disorder}, thus confirming the universality of stationary measures at large scale.   
\\

To obtain more detailed information, let us compute the cumulants of $X(L)$ for any $v$, on the line $u+v=-1$.
Using the general formula \eqref{cumgenY} for $w=u+v=-1$,  we obtain 
\be
\langle Y^n \rangle_L^c 
= \left(-\tfrac{1}{2} \partial_v\right)^n \left(v^2 L + \log(e^{(1+2 v) L}-1) - 
\log(1+ 2 v) \right) 
\ee 
where in the last equation we used \eqref{Norm1}. %Note that the moments are instead given by $\langle Y^n \rangle_L = [{\cal Z}_{-1-v,v,L}]^{-1} (-\frac{1}{2} \partial_v)^n {\cal Z}_{-1-v,v,L}$.
One finds for the mean
\bea \label{meanY} 
\langle Y \rangle_L = L \left(\frac{1}{1-e^{(2 v+1)L}}-v-1\right)+\frac{1}{2 v+1}
\eea 
The small $L$ behavior is $\langle Y \rangle  = - (v+ \frac{1}{2}) L + O(L^2)$. At large $L$ there are two cases
(i) $v > -1/2$ and $\langle Y \rangle  \simeq u L$ (using $u+v=-1$) and (ii) $v < -1/2$ and $\langle Y \rangle  \simeq -v  L$. 
Around the symmetric point $v=u=-1/2$ the mean vanishes as $\langle Y \rangle = - \frac{1}{6} L (6+L) (v+\frac{1}{2}) + O((v+\frac{1}{2})^2)$.
%\red{P: this seems consistent with what we know for you?} 

The variance is 
\bea 
\langle Y^2 \rangle^c_L = \frac{L^2}{2-2 \cosh ((2 v+1) L)}+\frac{L}{2}+\frac{1}{(2 v+1)^2}
\eea 
The small $L$ behavior is $\langle Y^2 \rangle  = \frac{L}{2} + O(L^2)$. For $v \neq -1/2$
the large $L$ behavior is simply $\langle Y^2 \rangle  \simeq L/2$. Exactly at $v=-1/2$, the
variance goes to a constant of order $L^2$, more precisely for any $L$ around the symmetric point one has 
$\langle Y^2 \rangle = \frac{1}{12} L (L+6)-\frac{1}{60} L^4
 (v+\frac{1}{2})^2+O\left((v+\frac{1}{2})^3\right)$. One sees from these results that at large $L$ 
and near the symmetric point, there is a diverging length scale $L_v =1/(v+ \frac{1}{2})$,
and a scaling behavior with $L/L_v$ (it will be studied in more details below). These results are consistent with the observation that
from \eqref{PX}, $P_L(Y)$ at large $L$ is approximately $e^{-(1+ 2 v) Y}$ times a 
uniform distribution on the interval $[-L/2,L/2]$.
\bigskip 

For $u+v=-1$ one can also easily obtain the one point PDF of $X(x)$, denoted $P_{x,L}(Y)$ for any $0<x<L$. By similar manipulations one finds
\be 
P_{x,L}(Y) = P_x(Y) \frac{{\cal Z}_{u,v,x} }{{\cal Z}_{u,v,L}} e^{v^2 (L-x)} + e^{-2 Y} \frac{e^{- \frac{(Y + v x)^2}{x}} }{\sqrt{\pi x}} \frac{{\cal Z}_{u,v,L-x} }{{\cal Z}_{u,v,L}} e^{v^2 x},
\ee 
where $P_x(Y)$ is obtained from $P_L(Y)$ in \eqref{PX} by substituting $L \to x$, and ${\cal Z}_{u,v,L}$ is given in \eqref{Norm1}. It is then 
easy to compute its mean using the result \eqref{meanY} and one thus finds {\it the mean profile} as 
\bea \label{meanYx} 
\mathbb{E}[ X(x) ] =  \langle Y \rangle_{x,L} &=& \langle Y \rangle_{x} \frac{{\cal Z}_{u,v,x} }{{\cal Z}_{u,v,L}} e^{v^2 (L-x)} - (v+1) x e^{2 v x+x} \frac{{\cal Z}_{u,v,L-x} }{{\cal Z}_{u,v,L}} e^{v^2 x}, \\
&=& \frac{v (2 v+1) x+e^{(2 v+1)x}-1 -(v+1) (2 v+1) x e^{(2 v+1) L}}{(2
   v+1) \left(e^{(2v+1) L}-1\right)}.
\eea 
At the symmetric point $v=u=-1/2$ the mean profile is parabolic, i.e. $\langle Y \rangle_{x,L} = - \frac{x(L-x)}{2 L}$.
This is consistent with the scenario discussed above. 
Near the symmetric point, and for large $L$, there is a critical regime with a diverging length scale $L_v = 1/(v+ \frac{1}{2})$
and the mean profile takes the scaling form
\be 
\langle Y \rangle_{x,L} \simeq \frac{L}{4}  \left(\frac{\left(e^{2 {\tilde v}  {\tilde x}}-1\right) (\coth
   ({\tilde v} )-1)}{{\tilde v} }-2 {\tilde x} \coth ({\tilde v} )\right), \quad \tilde x= \frac{x}{L}  , \quad \tilde v= \frac{L}{L_v}=L(v+\frac{1}{2}).
\ee 
\textbf{Remark: } A similar diverging length scale near $u=v<0$ was obtained in previous works on TASEP \cite{schutz1993time, krug1994disorder}, with the same power law dependence with respect to $u-v$. 
\\

\textbf{Case $u+v=-n$. } This calculation can be extended to the case $u+v=-n$ for any positive integer $n$, using that
\be 
\mathbb{E}_{B(L)=Y} \left[ \left( \int_0^L dx e^{-2 B(x)}  \right)^{n} \right] = \int_0^L \dots \int_0^L dx_1 \cdots dx_n e^{-2 \frac{Y}{L} \sum_i x_i +  \sum_{i,j} 
\min(x_i,x_j)- \frac{x_i x_j}{L}}
\ee 
although the algebra becomes rapidly tedious. Instead, one can use in \cite[Prop. 5.3]{matsumoto2005exponential} 
and obtain
%There is a result for $u+v=-n$ and in the absence of drift, i.e for $v=0$ (Prop. 5.3 in \cite{matsumoto2005exponential}) 
%\be 
%\frac{1}{\sqrt{2 \pi t} } e^{- x^2/(2 t)} \mathbb{E}[ (\int_0^t ds e^{2 {\sf B}(s)})^n | {\sf B}(t)=x] = 
%\frac{e^{n x}}{n! (2 \pi t^3)^{1/2} } \int_{|x|}^{+\infty} dr r e^{-r^2/(2 t)} (\cosh(r)-\cosh(x))^n 
%\ee 
and obtain
\be \label{Pgen} 
P_L(Y) = \frac{1}{{\cal Z}_{u,v,L}} 
\frac{2^{n+1} e^{- n Y- 2 v Y}}{n! (\pi L^3)^{1/2} } \int_{|Y|}^{+\infty} dr \, r e^{-r^2/L} (\cosh(r)-\cosh(Y))^n 
\ee 
where ${\cal Z}_{u,v,L}$ is given in \eqref{normN}. We can give a more explicit form for $u+v=-2$
\begin{multline}  \label{PX2} 
 P_L(Y) = \frac{e^{\frac{L}{4}-2 (v+1) Y}}{2 {\cal Z}_{u,v}} 
 \bigg(e^{\frac{3 L}{4}}
   \left(\text{erf}\left(\frac{L-Y}{\sqrt{L}}\right)+\text{erf}\left(\frac
   {L+Y}{\sqrt{L}}\right)\right) \\
    -2 \cosh Y
   \left(\text{erf}\left(\frac{L-2 Y}{2
   \sqrt{L}}\right)+\text{erf}\left(\frac{L+2 Y}{2
   \sqrt{L}}\right)\right)\bigg),
\end{multline}
where the normalization is
\be 
{\cal Z}_{u,v,L} = e^{v^2 L} \frac{2 v+3-4 (v+1) e^{2 L v+L}+(2 v+1) e^{4 L (v+1)}}{2 (v+1) (2 v+1) (2v+3)}.
\ee 

Returning to arbitrary integer $n$, we see that at large $L$ the integrand in \eqref{Pgen} is dominated by large $r$, where it behaves as $ e^{ - r^2/L + n r}$. 
Hence if $|Y| < L n/2$ the saddle point at $r=L n/2$ dominates and the integral is of order $ e^{L/4 n^2}$, i.e approximately independent of $Y$.
Hence for $u=v=-n/2$ the result is similar to the one obtained above for $n=1$. The exponential factor
$e^{- n Y- 2 v Y} = e^{(u-v) Y} $ will deform the distribution of $Y$ for $u-v \neq 0$ which becomes peaked
near $Y= {\rm sgn}(u-v) n L/2$.
\\

\textbf{Case $u+v=1$.} 
Another case where simplifications occur is $u+v=1$ where using  \cite[Prop. 5.9]{matsumoto2005exponential} and \eqref{54} one finds
\be 
P_L(Y) = \frac{1}{{\cal Z}_{u,v,L}} \frac{e^{ - \frac{Y^2}{L} -2 v Y}}{\sqrt{\pi L}} \frac{Y e^{Y}}{L \sinh Y} 
\ee 
and ${\cal Z}_{u,v,L}$ can be simply obtained from the normalization condition $\int_{\mathbb{R}} dY P_L(Y)=1$.
\\

\textbf{Case $u,v>0$. } In that case, we use the Laplace transform, obtained for $L=1$ in \cite{corwin2021stationary},  and which reads for general $L$ and 
for $-2 v < c < 2 u$
\be 
\mathbb{E}[ e^{-c X(L)} ] = \frac{I(c)}{I(0)} , \quad I(c)= \int_0^{+\infty} dk e^{- \frac{k^2}{4} L} \frac{ |\Gamma(\frac{c}{2} + v + \frac{i k}{2})|^2 |\Gamma(-\frac{c}{2} + u + \frac{i k}{2})|^2 }{|\Gamma(2 i k)|^2}.
\ee 
\\
For $L \to +\infty$ without rescaling $u,v>0$ one finds, by rescaling $k \to k/\sqrt{L}$
\be 
\mathbb{E}[ e^{-c X(L)} ] \xrightarrow[L\to+\infty]{}\frac{\Gamma(\frac{c}{2} + v)^2 \Gamma(-\frac{c}{2} + u)^2 }{\Gamma(v)^2 \Gamma(u)^2}
\ee 
Since a Gamma random variable $\gamma_u$ of parameter $u$ has moments $\mathbb{E}[ \gamma_u^a ] = \frac{\Gamma(u+a)}{\Gamma(u)}$, we obtain the weak convergence 
\be 
X(L) \Longrightarrow - \frac{1}{2} \log \left( \frac{\gamma_v^{(1)} \gamma_v^{(2)}}{\gamma_u^{(1)} \gamma_u^{(2)} } \right) , 
\ee 
where the four Gamma random variables are independent. Let us recall that the total height difference is
$H(L) - H(0) = \sqrt{\frac{L}{2}} G + X(L)$. Since $X(L)$ is $O(1)$ at large $L$ it is subdominant in the variance of $H(L)$,
but it controls the higher cumulants of the height difference.
Note that $X(L)=O(1)$ is consistent with the rescaled process 
$\widetilde X(\tilde x)=\frac{1}{\sqrt{L}} X(L \tilde x)$ being a Brownian excursion in the large $L$ limit, as shown in the Letter.

\section{Some properties of the steady state of the KPZ fixed point on the interval}

\subsection{Multipoint probability for the rescaled process, Brownian motion with an absorbing boundary, and quantum mechanics in presence of a hard wall} 
We have shown in the Letter that, in the limit $L \to +\infty$, upon rescaling the boundary parameters
as $u=\tilde u/\sqrt{L}$, $v=\tilde v/\sqrt{L}$, and rescaling space as $\tilde x=x/L$, the rescaled 
process $\widetilde X(\tilde x) = \frac{1}{\sqrt{L}} X(\tilde x)$, $\tilde x \in [0,1]$
has the following probability measure, which can be written in the two equivalent forms
\be \label{measurerescaled} 
\frac{e^{-\tilde v^2}}{\widetilde Z_{\tilde u,\tilde v}} \mathcal D \widetilde X e^{ -  \int_0^1  d\tilde x  \left(\frac{d\tilde X(\tilde x)}{d\tilde x}+ \tilde v\right)^2  } 
e^{2  (\tilde u + \tilde v) \min_{\tilde x} \widetilde X(\tilde x)   } = \frac{1}{\widetilde Z_{\tilde u,\tilde v}} 
\mathcal D \widetilde X e^{ -  \int_0^1  d\tilde x  (\frac{d\tilde X(\tilde x)}{d\tilde x})^2  }
e^{2  (\tilde u + \tilde v) \min_{\tilde x} \widetilde X(\tilde x) - 2 v \widetilde X(1)  }
\ee 
with $\widetilde X(0)=0$ and where $\widetilde Z_{\tilde u,\tilde v}$ is a normalisation. Consider an observable $O\left[\widetilde X\right]$. Its expectation is given by
\bea \label{2lines} 
 \mathbb{E}_{\widetilde X}\left[ O[\widetilde X] \right] &=& \frac{e^{-\tilde v^2}}{\widetilde Z_{\tilde u,\tilde v}} \int_{-\infty}^0 db  e^{2  (\tilde u + \tilde v) b }  
 \mathbb{E}_{B_{-\tilde v}}\left[ O[B_{-\tilde v}] \delta( \min B_{-\tilde v} - b ) \right],\\
&=& \frac{1}{\widetilde Z_{\tilde u,\tilde v}} \int_{-\infty}^0 db e^{2  (\tilde u + \tilde v) b }  \mathbb{E}_B\left[ O[B] e^{-2 \tilde v B(1)} \delta( \min B - b) \right],
\eea 
where the averages are over a Brownian motion of diffusion coefficient $1/2$, respectively with drift $- \tilde v$ and with zero drift, on the interval $[0,1]$. We denote the minimum of the process $B$ over $[0,1]$ by $\min B$.

Let us consider $\tilde u+\tilde v>0$ and perform an integration by part
\bea 
\mathbb{E}_{\widetilde X}\left[ O[\widetilde X] \right] &=& \frac{e^{-\tilde v^2}}{\widetilde Z_{\tilde u,\tilde v}} 2  (\tilde u + \tilde v) \int_{-\infty}^0 db 
 e^{2  (\tilde u + \tilde v) b }  \mathbb{E}_{B_{-\tilde v}}\left[ O[B_{-\tilde v}] \theta(\min B_{-\tilde v} - b) \right] \\
 &=& \frac{1}{\widetilde Z_{\tilde u,\tilde v}} 2  (\tilde u + \tilde v) \int_{-\infty}^0 db 
 e^{2  (\tilde u + \tilde v) b }  \mathbb{E}_{B}\left[ O[B] e^{-2 \tilde v B(1)} \theta( \min B - b ) \right]
 \label{eq:computingobservables}
\eea 

For a Brownian motion it is easy to evaluate the multi-point probability joint with the probability that the minimum is larger than $b$. Let us introduce the 
propagator of the Brownian motion in presence of an absorbing wall at position $U=b$, which can be obtained from the image method
\be 
G_{b}(U,U',x-x') = G(U-b,U'-b,x-x'), \quad G(U,U',x-x') = \left( \frac{e^{- \frac{(U-U')^2}{x-x'}} - e^{- \frac{(U+U')^2}{x-x'}}}{\sqrt{\pi (x-x')} } \right) \theta(U) \theta(U')
\ee
One has
\be 
G_{b}(U,U',x-x') = \mathbb{E}_{B}\left[\left. \delta(B(x)- U) \theta\left( \min_{x'<z<x} B(z) - b \right) \right\vert B(x')= U' \right] \theta(U'-b).
\ee 
For the propagator with drift, it is similar.

The evaluation of the multi-point probabilities can be performed equivalently using the quantum mechanics of a particle in presence of a hard wall. Indeed one can also write
\be  \label{sinus} 
G(U,U',x-x') =  \int_0^{+\infty}  dk \phi_k(U) \phi_k(U') e^{- (x-x') \frac{k^2}{4}}, \quad \phi_k(U)=\sqrt{ \frac{2}{\pi} } \sin(k U) \,  \theta(U) 
\ee 
This is the Green's function for the Hamiltonian given by Eq. (8) in the Letter, where the Liouville potential $V(U)=e^{2 U}$ is replaced by a hard wall
potential, i.e. $V(U)=+\infty$ for $U<0$ and $V(U)=0$ for $U>0$, which has eigenfunctions $\phi_k$. It is natural that the hard wall potential arises
in the large $L$ limit, since the rescaling of the process $X$ leads to $U \to \sqrt{L} U$. 
The formula for this quantum mechanics can be obtained by considering all LQM formula in the Letter in the regime of
small $k$. One can check that in this limit the LQM eigenfunctions $\psi_k(U)$ given in Eq. (9) in the Letter become equal to the
eigenfunction $\phi_k(U)$ in \eqref{sinus} restricted to $U>0$, upon rescaling $k \to k/\sqrt{L}$ and $U \to \sqrt{L} U$.
\\

Let us return to the computation of the multipoint distribution of the process $\widetilde X$. 
For $\tilde x_0=0<\tilde x_1<\dots<\tilde x_m<\tilde x_{m+1}=1$, $ b_1, \dots, b_{m+1} \in \mathbb R$, $b\leqslant 0$ and $b_0=0$, we may write 
\begin{multline}
 e^{- \tilde v^2} \mathbb{E}_{B_{-\tilde v}} \left[ \delta(B_{-\tilde v}(\tilde x_1) - b_1) \dots \delta(B_{-\tilde v}(\tilde x_m)-b_m) \delta(B_{-\tilde v}(1)-b_{m+1}) \theta( \min B_{-\tilde v} - b) \right] \\
 = \mathbb{E}_{B} \left[ \delta(B(\tilde x_1) - b_1) \dots \delta(B(\tilde x_m)-b_m) \delta(B(1)-b_{m+1}) e^{-2 \tilde v B(1)} \theta( B - b) \right] \\
 = e^{ - 2 \tilde v b_{m+1}} \prod_{j=0}^m G_{b}(b_{j+1},b_j,\tilde x_{j+1}-\tilde x_j). 
\end{multline}
From this result,  one obtains using \eqref{eq:computingobservables} the multipoint PDF of the rescaled process $\widetilde X$ (for $\tilde u + \tilde v>0$) 
\begin{multline} \label{multi1} 
 \mathbb{E}_{\widetilde X} \left[ \delta(\widetilde X(\tilde x_1) - \widetilde X_1) \dots \delta(\widetilde X(\tilde x_m)- \widetilde X_m) \delta(\widetilde X(1)- \widetilde X_{m+1}) \right]
\\
 = \frac{1}{\widetilde Z_{\tilde u,\tilde v}} 2  (\tilde u + \tilde v) \int_{-\infty}^0 db e^{2  (\tilde u + \tilde v) b } 
 e^{ - 2 \tilde v \widetilde X_{m+1}} \prod_{j=0}^m G_{b}(\widetilde X_{j+1},\widetilde X_j, x_{j+1}-x_j)
\end{multline}
This formula is suitable to calculate the Laplace transform in the form
\be 
\mathbb{E}_{\widetilde X} \left[ e^{- \sum_{j=1}^m \tilde s_j (\widetilde X(\tilde x_j)-\widetilde X(\tilde x_{j+1})) } \right] = \frac{\hat J\left(\vec{\tilde s}\right)}{\hat J(0)}
\ee
From similar manipulations as in the Letter (from (16) to (20)) but using the eigenfunctions for the hard wall quantum mechanics \eqref{sinus}, 
it is possible to express it as 
\begin{multline}  \label{eq:finalJFP} 
\hat J\left(\vec{\tilde s}\right) =    \prod_{j=1}^{m+1} \int_0^{+\infty}  \frac{dk_j k_j^2}{4 \pi} 
\prod_{j=1}^m \frac{(\tilde s_j - \tilde s_{j+1})}{((\tilde s_j - \tilde s_{j+1})^2 + (k_j+k_{j+1})^2 ) ((\tilde s_j - \tilde s_{j+1})^2 + (k_j-k_{j+1})^2 ) }\\
 \times \frac{1}{(2 \tilde u- s_1)^2 + k_1^2} \frac{1}{4 \tilde v^2 + k_{m+1}^2} \, e^{ - \sum_{j=1}^{m+1}  \frac{k_j^2}{4}(x_j-x_{j-1}) },
\end{multline}
where $2\tilde u>\tilde s_1>\dots>\tilde s_{m+1}$. 
This formula  coincides, up to an irrelevant global multiplicative constant, with the formula (20) in the Letter, 
where one performs the rescaling $s_j \to \tilde s_j/\sqrt{L}$, $k_j \to \tilde k_j/\sqrt{L}$, $x_j \to L \tilde x_j$,
$u \to \tilde u/\sqrt{L}$, $v \to \tilde v/\sqrt{L}$. 
\\

Finally, note that for arbitrary $\tilde u + \tilde v$ we can obtain the multipoint joint PDF of the rescaled process $\widetilde X$ in the form 
\begin{multline} 
 \mathbb{E}_{\widetilde X} \left[ \delta(\widetilde X(\tilde x_1) - \widetilde X_1) \dots \delta(\widetilde X(\tilde x_m)- \widetilde X_m) \delta(\widetilde X(1)- \widetilde X_{m+1}) \right]
\\
 = \frac{1}{\widetilde Z_{\tilde u,\tilde v}} \int_{-\infty}^0 db e^{2  (\tilde u + \tilde v) b } 
 e^{ - 2 \tilde v \widetilde X_{m+1}} (-\partial_b) \left[ \prod_{j=0}^m G_{b}(\widetilde X_{j+1},\widetilde X_j, \tilde x_{j+1}-\tilde x_j) \right]
\end{multline}

\subsection{Normalisation}

It is easy to calculate the normalisation of the measure of the rescaled process \eqref{measurerescaled} using
known results about the maximum of the Brownian motion. One finds
\be  \label{normresc} 
\widetilde Z_{\tilde u,\tilde v} = \int_{-\infty}^0 dy \int_{y}^{+\infty} dY 
\frac{4 (Y-2 y)}{\sqrt{\pi} } e^{- (2 y-Y)^2} 
e^{2 (\tilde u + \tilde v) y - 2 \tilde v Y}
= \frac{\tilde u e^{\tilde u^2}  \text{erfc}(\tilde u)- \tilde v e^{\tilde v^2}  \text{erfc}(\tilde v)}{\tilde u-\tilde v}
\ee 
where we have used that the joint PDF of $\min B=y$ and $B(1)=Y$
is $2 p(- \sqrt{2} y, - \sqrt{2} Y)$ where $p$ is given in (1.9) in \cite{shepp1979joint}.
It can also be obtained by Laplace inversion from \cite[Chap. IV, item 32]{borodin2015handbook}.
\\
%the joint PDF of the max and endpoint of a standard Brownian is
%\be 
%p(y,x)=  \frac{2 (2 y-x)}{\sqrt{2 \pi} } e^{- \frac{1}{2} (2 y-x)^2} \theta(y) \theta(y-x) 
%\ee 
%\be
%2 p(- \sqrt{2} y, - \sqrt{2} x) =  \frac{4 (x-2 y)}{\sqrt{\pi} } e^{- (2 y-x)^2} \theta(-y) \theta(x-y) 
%\ee 

\subsection{One point distribution of $\widetilde X(1)$} 

The formula \eqref{multi1} with $m=0$ gives for $\tilde u+\tilde v>0$ the PDF of the random variable $\widetilde X(1)$ as
\be \label{multi1bis}
P(Y) = \mathbb{E}_{\widetilde X} \left[  \delta(\widetilde X(1)- Y) \right]  = \frac{1}{\widetilde Z_{\tilde u,\tilde v}} 2  (\tilde u + \tilde v) e^{ - 2 \tilde v Y} \int_{-\infty}^0 db e^{2  (\tilde u + \tilde v) b } 
\frac{1}{\sqrt{\pi}} \left( e^{- Y^2} - e^{- (Y-2 b)^2} \right) \theta(Y-b)
\ee
which leads to 
\be \label{PYresc} 
P(Y)  = \frac{2  (\tilde u + \tilde v)}{\widetilde Z_{\tilde u,\tilde v}}  e^{ - 2 \tilde v Y} \left(  \frac{e^{-Y^2}}{2 \sqrt{\pi }
   (\tilde u+\tilde v)} (\theta(Y) + \theta(-Y) e^{2 Y (\tilde u+\tilde v)} ) - \frac{1}{4} e^{\frac{1}{4} (\tilde u+\tilde v) (\tilde u+\tilde v+4 Y)}
   \text{erfc}\left(\frac{\tilde u+\tilde v}{2}+|Y| \right) \right)
\ee 
Using the normalisation in \eqref{normresc}, one checks that it is correctly normalized to unity $\int_{\mathbb{R}} dY P(Y)=1$.
\\

{\bf Laplace transform}. Alternatively, for $\tilde u+\tilde v>0$ we can use the Laplace transform, obtained for $L=1$ in \cite{corwin2021stationary}, extended here
to general $L$, and which leads, for the rescaled process in the limit $L \to +\infty$ (as discussed in the previous section
one needs to rescale $k \to k/\sqrt{L}$) to the formula
\be 
\mathbb{E}[ e^{-\tilde c \widetilde X(1)} ] = \frac{I(\tilde c)}{I(0)}  , \quad I(\tilde c)= F(2 \tilde v+ \tilde c,2 \tilde u - \tilde c) , \quad 
F(a,b)=
\int_0^{+\infty} dk e^{- \frac{k^2}{4} } 
\frac{k^2}{(k^2 + a^2) (k^2 + b^2)} 
\ee 
One has
\be 
F(a,b)= \frac{a^2 g(a) - b^2 g(b)}{a^2-b^2}  , \quad g(a) = \int_0^{+\infty} dk e^{- \frac{k^2}{4} } \frac{1}{k^2 + a^2}
= \frac{\pi  e^{\frac{a^2}{4}} \text{erfc}\left(\frac{a}{2}\right)}{2 a}
\ee 
This leads to the formula, for $c\in (-2\tilde v, 2\tilde u)$, 
\be 
\mathbb{E}\left[ e^{- c \widetilde X(1)} \right] = \frac{e^{\frac{1}{4} c (c-4 \tilde u)} (\tilde v-\tilde u) \left(e^{\tilde u^2} (c-2 \tilde u)
   \text{erfc}\left(\tilde u-\frac{c}{2}\right)+(c+2 \tilde v)
   \text{erfc}\left(\frac{c}{2}+\tilde v\right) e^{c (\tilde u+\tilde v)+\tilde v^2}\right)}{(c-\tilde u+\tilde v)
   \left(2 e^{\tilde v^2} \tilde v \, \text{erfc}(\tilde v)-2 e^{\tilde u^2} \tilde u \,\text{erfc}(\tilde u)\right)}
\ee
Expanding in $c$, this formula leads to expressions for the mean and variance of $\widetilde X(1)$. 
The Laplace inversion of that formula is not a priori obvious. 
One can check however (through a tedious calculation) that it is indeed the Laplace transform of $P(Y)$ in \eqref{PYresc}. 
This shows the consistency with the claims of the previous section, and that the method using the Brownian representation of the measure
is quite efficient to obtain the PDF's of the process. 
\\

\subsection{Properties of the minimum of the rescaled process $\widetilde X$} 

Let $\widetilde X_m$  be the minimum of $\widetilde X(\tilde x)$ on the interval $\tilde x \in [0,1]$ and let  $0 \leq \tilde x_m \leq 1$ be the first point where it is reached. Then the joint
PDF of $\widetilde X_m$, $\tilde x_m$ and of the value $Y=\widetilde X(1)$ of the process at the boundary  is given by 
\be \label{max1} 
P(\widetilde X_m,\tilde x_m,Y) = C_{u,v} P^B(\widetilde X_m,\tilde x_m,Y) e^{2 (\tilde u + \tilde v) \widetilde X_m- 2 \tilde v Y }, 
\ee 
where $C_{u,v}$ is a normalization, and $P^B(\widetilde X_m,x_m,Y)$ is the corresponding joint PDF for the
Brownian $B$ with diffusion coefficient $1/2$ and zero drift. Using the main result of  \cite{shepp1979joint}
we obtain
\be 
P(y,x,Y) = \frac{1}{{\cal Z}_{u,v}} \frac{4 |y| (Y-y)}{\pi x^{3/2} (1- x)^{3/2} } e^{- \frac{y^2}{x} - \frac{(y-Y)^2}{1- x} + 2 (\tilde u + \tilde v) y - 2 \tilde v Y} \theta(-y) \theta(Y-y) 
\theta(0 < x<1)
\ee 
Integration over $x$, the position of the minimum, using the identity below (1.9) in \cite{shepp1979joint} leads to the joint
PDF of the minimum and the value at the boundary as
\be 
P(y,Y) = \frac{1}{{\cal Z}_{u,v}} \frac{4 (Y-2 y)}{\sqrt{\pi} } e^{- (Y-2 y)^2+  2 (\tilde u + \tilde v) y - 2 \tilde v Y} \theta(-y) \theta(Y-y) 
\ee 
This PDF is correctly normalized to unity, from \eqref{normresc}. Integrating over $y$ we obtain once again (by a different method)
the result \eqref{PYresc}. However, with this method it appears to be valid for any $\tilde u$,  $\tilde v$.

\section{Translation of the results of Hariya-Yor}

\begin{figure}
    \centering
    \begin{tikzpicture}[scale=1, every text node part/.style={align=center}]
    %\fill[blue,opacity=0.1] (0,0) --(2,2) -- (2,4) -- (0,4) -- cycle;
    %\fill[green,opacity=0.15] (0,0) --(2,2) -- (4,2) -- (4,0) -- cycle;
    %\fill[yellow,opacity=0.2] (2,2) --(2,4) -- (4,4) -- (4,2) -- cycle;
%    \draw[thick, black!20] (0,4) -- (4,0);
    \draw[thick] (0,0) -- (2,2);
    \draw[thick, ->] (2,2) -- (2,4.1) node[anchor=north east] {$v$};
    \draw[thick, ->] (2,2) -- (4.1,2) node[anchor=north] {$u$};
    \draw[dashed, gray] (2,0) -- (2,2);
    \draw[dashed, gray] (0,2) -- (2,2);
    \draw[gray] (2,0.1) -- (2,-0.1) node[anchor=north] {$0$};
    \draw[gray] (0.1,2) -- (-0.1,2) node[anchor=east] {$0$};
    \draw (1,2.5) node{$R_2$};
    \draw (2.5,1) node{$R_3$}; 
    \draw (3,3) node{$R_1$}; 
    \draw (1,1) node{$L_2$}; 
    \draw (3,2) node{$L_3$};  
    \draw (2, 3) node{$L_1$};  
    \end{tikzpicture}
    \caption{The regions $R_1,R_2,R_3$ and their borders $L_1,L_2,L_3$  defined in \cite{hariya2004limiting}.} 
    \label{fig:HaryiaYor}
\end{figure}
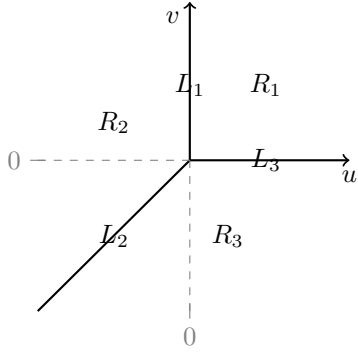

Let us state the results of  \cite{hariya2004limiting,matsumoto2005exponential} and translate them into our notations. In the Letter, we consider the measure on continuous real-valued processes $X(x) = \hat X(x) - v x$, defined on on $[0,L]$, such that the density of $\hat X$ is 
\be 
\mathcal D \hat X e^{- \int_0^L dx (\frac{d \hat X}{dx})^2 } \left(\int_0^L dx e^{-2 \hat X(x) + 2 v x}\right)^{- (u+v)}. 
\ee 
Hariya and Yor \cite{hariya2004limiting} studied several probability measures on Brownian paths weighted by exponential functionals of the Brownian motion. In particular, they study the measure on processes $X(s)={\sf B}(s)+\mu s$ defined on $[0,t]$, denoted $\mathcal P_t^{(\mu,m)}$ in \cite{hariya2004limiting}, defined by 
\be
\mathcal D {\sf B} e^{- \frac{1}{2} \int_0^t ds (\frac{d{\sf B}}{ds})^2 }  \left(\int_0^t ds e^{2 {\sf B}(s) + 2 \mu s}\right)^{-m}. 
\ee 
where ${\sf B}$ denotes the standard Brownian motion, to distinguish it from $B$ which we reserve for the Brownian with diffusion coefficient $1/2$, i.e.
$B= \frac{1}{\sqrt{2}} {\sf B}$.
In order to match the measures, we let $x=2 s$,  that is
\be \label{correspondence1} 
\hat X(x) = - B(s=x/2). 
\ee 
Then our measure becomes 
\be 
\mathcal D {\sf B} e^{- \frac{1}{2} \int_0^{L/2} ds (\frac{d {\sf B}}{ds})^2 } \left(\int_0^{L/2} ds e^{2 {\sf B}(s) + 4 v s}\right)^{- (u+v)}. 
\ee 
Hence we must identify
\be \label{correspondence2} 
m = u+ v,  \quad \mu = 2 v  , \quad t =L/2. 
\ee 
Haryia-Yor \cite{hariya2004limiting}  define three regions (see Fig. \ref{fig:HaryiaYor}), which exactly match the three regions of our phase diagram:  
\begin{align} 
 R_1&= \{ 2 m > \mu , \mu>0 \} = \{u>0,v >0 \}  \\
 R_2&= \{ m < \mu , 2 m < \mu \} = \{u<v, u<0 \}  \\
R_3 &= \{ m > \mu ,  \mu <0 \} = \{u>v, v<0 \}  
\end{align} 
They also defined the borders between these regions
\begin{align}  
 L_1 &= \{ 2 m = \mu , \mu>0 \} = \{u=0,v >0 \},\\
 L_2&= \{ m = \mu , \mu<0 \} = \{u=v,v <0 \}, \\
 L_3 &= \{ m >0  , \mu=0 \} = \{u>0,v=0 \}, 
\end{align} 
so that $L_1$ is between $R_1$ and $R_2$, $L_2$ is between $R_2$ and $R_3$, $L_3$ is between $R_1$ and $R_3$. 
\medskip 

Hariya and Yor proved  \cite[Theorem 1.3]{hariya2004limiting} (the statement can also be found in the review \cite[Theorem 7.4]{matsumoto2005exponential}) that we have the following weak convergence of probability measures on continuous functions: 
\begin{equation}
\mathcal P_t^{(\mu,m)} \xRightarrow[t\to+\infty]{} \begin{cases} {\cal W}^{(0)}_{2 \gamma_{\nu=m- \mu/2} } = {\cal W}^{(0)}_{2 \gamma_{\nu=u} } &\mbox{ if } (u,v)\in L_1\cup R_1 \cup  L_3, \\
 {\cal W}^{(\mu-2 m)} = {\cal W}^{(-2 u)} &\mbox{ if }(u,v)\in L_1 \cup R_2 \cup L_2,\\
 {\cal W}^{(-\mu)}_{2 \gamma_{\nu=m-\mu}} = {\cal W}^{(-2 v)}_{2 \gamma_{u-v}} &\mbox{ if }(u,v)\in L_2\cup R_3 \cup L_3.
 \end{cases} 
 \label{eq:HariyaYorTheorem}
\end{equation}
%\bea 
% \mbox{ For }(u,v)\in L_1\cup R_1 \cup  L_3 :  &\mathcal P_t^{(\mu,m)} &\Longrightarrow {\cal W}^{(0)}_{2 \gamma_{\nu=m- \mu/2} } = {\cal W}^{(0)}_{2 \gamma_{\nu=u} } ,\\
%\mbox{ For }(u,v)\in L_1 \cup R_2 \cup L_2 :  & \mathcal P_t^{(\mu,m)} &\Longrightarrow {\cal W}^{(\mu-2 m)} = {\cal W}^{(-2 u)} ,\\
% \mbox{ For }(u,v)\in L_2\cup R_3 \cup L_3  :  &\mathcal P_t^{(\mu,m)} &\Longrightarrow {\cal W}^{(-\mu)}_{2 \gamma_{\nu=m-\mu}} = {\cal W}^{(-2 v)}_{2 \gamma_{\nu=u-v}},
%\eea 
In \eqref{eq:HariyaYorTheorem}, $\gamma_\nu$ is a Gamma random variable with parameter $\nu$ (and by convention, $\gamma_0=0$), ${\mathcal W}^{(\mu)}$ denotes  the law of a Brownian motion with drift $\mu$ , and ${\mathcal W}_{2\gamma_{\nu}}^{(\mu)}$ denotes the law of the process 
\begin{equation} t \mapsto {\sf B}_t+\mu t -\log\left(  1+ 2\gamma_{\nu} \int_0^{t} e^{2 {\sf B}_s+2\mu s}ds\right).\end{equation}
{Using \eqref{correspondence1} this implies that the process $X(x)$ converges as $L$ goes to infinity to the following, consistent with  Equations (34) and (35) in the Letter. 
\begin{itemize} 
\item When $u,v\geqslant 0$,  $X(x)$ weakly  converges to 
\begin{equation} 
 B(x)+\log\left( 1+\gamma_u \int_0^x e^{-2B(z)}dz\right).
 \label{eq:HY1}
\end{equation}
\item When $u\leqslant 0$ and $u\leqslant v$, $X(x)$ weakly converges to
\begin{equation}
B(x) +ux.
\label{eq:HY2}
\end{equation} 
\item And, when $v\leqslant 0$, $u\geqslant v$, $X(x)$ weakly  converges to 
\begin{equation}
  B(x) + vx \\ +\log\left( 1+\gamma_{u-v} \int_0^x e^{-2B(z)-2vz}dz\right).
\label{eq:HY3}
\end{equation} 
\end{itemize} 
In \eqref{eq:HY1}, \eqref{eq:HY2} and \eqref{eq:HY3}, $B(x)$ denotes a Brownian motion with diffusion coefficient $1/2$, and $\gamma_u$ and $\gamma_{u-v}$ denote independent Gamma distributed random variables with respective shape parameters $u$ and $u-v$, and scale parameter $1$. 

\medskip 

When $u=-v\geqslant 0$, the process in \eqref{eq:HY3} becomes simply a Brownian motion with drift $u=-v$ and diffusion coefficient $1/2$. This non-trivial identity can be found in \cite[Eq. (1.8)]{hariya2004limiting}, see also \eqref{idelaw} below. }

\section{Review of the main arguments  in \cite{hariya2004limiting}} 

Here we describe the derivation of the convergence result \eqref{eq:HariyaYorTheorem}, that is  \cite[Theorem 1.3]{hariya2004limiting}, using the notation of the present paper. 

\subsection{Overall argument} Consider the process $X(x)$ with $0\leq x\leq L$. We are interested in taking $L \to +\infty$ and describing 
the process $X(z)$ for all $z \leq x$, where $x$ is fixed as $L\to +\infty$ . For this we can split the process
in two parts: $X(z)$ for $0\leq z \leq x$ and $X(z)$ for $x<z\leq L$. We will average over (i.e. integrate over) the second part of the
process, to obtain the measure for the first part. To characterize the measure, we consider  a general bounded functional $F_x[X]=F[\{X(z)\}_{0\leq z \leq x}]$ which depends only  on the first part. One has, from Equation (7) in the Letter 
\be 
\mathbb{E}_X\left[ F_x[X] \right] = \frac{1}{{\cal Z}_{u,v,L}} \mathbb{E}_{B_{-v}}\left[ F_x[B_{-v}] 
\left( \frac{1}{ \int_0^x dz e^{-2 B_{-v}(z)} + \int_x^L dz e^{-2 B_{-v}(z)} } \right)^{u+v}  \right] 
\ee 
where in the LHS the average is over the full process $\{X(z)\}_{0\leq z \leq L}$ and in the RHS 
it is over a Brownian motion with diffusion coefficient $1/2$ and drift $-v$, denoted $\{B_{-v}(z)\}_{0\leq z \leq L}$.
In the denominator we have splitted the integral into two pieces. Since the increments of the Brownian motion
after time $x$ are independent of the value $B_{-v}(x)$, we have the equality in law (for a fixed $x$)
\be
\int_x^L dz e^{-2 B_{-v}(z)}  = e^{-2 B_{-v}(x)} \int_x^L dz e^{-2 (B_{-v}(z)- B_{-v}(x)) } \overset{(d)}{=} e^{-2 B_{-v}(x) } \int_0^{L-x} dz e^{-2 \tilde B_{-v}(z)}
\ee 
where $\tilde  B_{-v}(z)$ is an independent Brownian motion with diffusion coefficient $1/2$ and drift $-v$. Hence we can 
rewrite 
\be  \label{average2bis} 
 \mathbb{E}_X[ F_x[X] ] =  \mathbb{E}_{\{ B_{-v}(z) \}_{0 \leq z \leq x}}\left[ F_x[B_{-v}] Q_x[B_{-v}] \right] 
\ee
in terms of the weight 
\be \label{average3} 
 Q_x[B_{-v}] =  \frac{\Delta(a,\xi, L-x)}{\Delta(0, 1, L)} , \quad a=\int_0^x dz e^{-2 B_{-v}(z)}  , \quad 
 \xi=e^{-2 B_{-v}(x)}
\ee
where we have defined 
\be 
\Delta(a, \xi, L) = \mathbb{E}_{\tilde B_{-v}} \left[ ( a + \xi Z_L )^{- (u+v)}  \right] , \quad Z_L = \int_0^{L} dz e^{-2 \tilde B_{-v}(z)} = \int_0^{L} dz e^{-2 \tilde B(z)+ 2 v z}, 
\ee 
which depends only on the random variable $Z_L$. 
\medskip 

The variable $\Delta(a, \xi, L)$ has been much studied and by \cite[Theorem 2.2]{hariya2004limiting}, translated 
in our variables using \eqref{correspondence1},\eqref{correspondence2}, we have 
\begin{equation} 
 \lim_{L \to +\infty} \frac{\Delta(a, \xi, L-x)}{\Delta(0,1, L)} = \begin{cases} 
 e^{v^2 x} \xi^{-v} \times \frac{1}{\Gamma(u)} \int_0^{+\infty} \frac{r^{u-1} e^{-r} }{(a r + \xi)^u} , & u>0,v \geq 0   \\
  e^{(v^2-u^2) x} \xi^{-(u+v)} , & u \leq 0 , u \leq v  \\
  \frac{1}{\Gamma(u-v)} \int_0^{+\infty} \frac{r^{u-v-1} e^{-r} }{(a r + \xi)^{u+v}}  , &  v \leq 0 , v<u
\end{cases} 
\label{limit1}
\end{equation} 
where the integrals arise as expectation values over independent Gamma variables, e.g. one has
\be 
\mathbb{E}_{\gamma_b}\left[\frac{1}{(\gamma_b a + \xi)^{c} }\right] = \frac{1}{\Gamma(b)} \int_0^{+\infty} dr  \frac{r^{b-1} e^{-r}}{ (a r  + \xi )^{c}  }
\ee 
where $\gamma_b$ denotes the Gamma variable $\Gamma(b,1)$. For instance, for $u>v$, the limit in the third line is easily understood since for $v<0$ one knows that
$\lim_{L \to +\infty} Z_L=Z_\infty$, where $1/Z_\infty$ is a Gamma variable $\Gamma(-2 v,1)$. For the other cases, see \cite[Theorem 2.2]{hariya2004limiting}.

\subsection{Large $L$ limits} Now we are able to take the large $L$ limits of \eqref{average2bis}. 
The simplest case to analyze is $u \leq 0 , u \leq v$. Inserting the limit in the second line of \eqref{limit1}, with $\xi=e^{-2 B_{-v}(x) }$, into \eqref{average2bis} one sees
that the measure for $X(z)$, $0 \leq z \leq x$ is 
\be 
\mathcal D X e^{(v^2-u^2) x} e^{- \int_0^x dz  \left(\frac{dX(z)}{dz} + v\right)^2 + 2 (u +v) X(x)  } = \mathcal D X e^{- \int_0^x dz \left( \frac{dX(z)}{dz} - u\right)^2  }
\ee 
with $X(0)=0$, 
so $X(z)$ is a Brownian with diffusion coefficient $1/2$ and drift $u$, as stated in the Letter. Note that the
multiplication by the weight $e^{2 (u +v) X(x)}$ has changed the drift from $-v$ to $u$. This is a general fact, {which is a special  case of the Cameron-Martin (CM) theorem}. 

\medskip 

Consider now the limit in the first line in \eqref{limit1}, that is when $u>0$, $v \geq 0$. Inserting it into \eqref{average2bis} and using the CM theorem (from the first line to the second) yields
\begin{align}
 \mathbb{E}_X[ F_x[X] ] &= \mathbb{E}_{\{ B_{-v}(z) \}_{0 \leq z \leq x}}\left[ F_x[B_{-v}] e^{v^2 x + 2 v B_{-v}(x)} 
\mathbb{E}_{\gamma_u}\left[ \frac{1}{(\gamma_u \int_0^x dz e^{-2 B_{-v}(z)}  + e^{-2 B_{-v}(x)} )^u }  \right] \right] \\
& = \mathbb{E}_{\{ B(z) \}_{0 \leq z \leq x}}\left[ F_x[B] 
\mathbb{E}_{\gamma_u}\left[ \frac{1}{(\gamma_u \int_0^x dz e^{-2 B(z)}  + e^{-2 B(x)} )^u }  \right] \right] \\
& = \mathbb{E}_{\{ Y(z) \}_{0 \leq z \leq x}}\left[ F_x[Y]  \right] \label{Y} 
\end{align} 
where in the second line $B$ has no drift. The non-trivial part is the third line, i.e. after expectation over $\gamma_u$ the average is
the same that over the path transformed process, 
\begin{equation} 
Y(z)=B(z) + \log\left( 1 + \gamma_u \int_0^z dz' e^{-2 B(z')} \right)
\end{equation} 
which gives the result stated in the {Letter (and above in \eqref{eq:HariyaYorTheorem})}. A similar identity allows to obtain the result stated {above in the third line in \eqref{eq:HariyaYorTheorem}, using  \eqref{limit1}, when $v<u$, $v \leq 0$}.

\subsection{The $T$-transformation on paths}
For completeness, we briefly recall here how the identity \eqref{Y} is derived in \cite{hariya2004limiting}. For simplicity we will
return to ${\sf B}_\mu$, the standard Brownian motion with drift $\mu$ used there, and use the correspondence \eqref{correspondence1},\eqref{correspondence2}
at the end. The argument is based on: 
\begin{enumerate} 
\item[(i)] The definition of the path transformation $T_z$ (also used in \cite{donati2001some}) defined for any process $X$ as
\be 
T_z(X)(t) = X_t - \log\left(1 + z \int_0^t ds e^{2 X(s)}\right) 
\ee 
which has the following interesting properties 
\be 
\frac{1}{\int_0^t ds e^{2 T_z(X)(s) } } = \frac{1}{\int_0^t ds e^{2 X(s) } } + z , \quad T_{z} T_{z'} = T_{z+z'} , \quad 
e^{X(t)} = \frac{e^{T_z(X)(t)}}{1- z \int_0^t ds e^{2 T_z(X)(s)}} \label{identities} 
\ee 
The first follows from the total derivative property 
\begin{equation} e^{2 T_z(X)(t)} = \frac{e^{2 X(t)}}{(1+ z \int_0^t ds e^{2 X(s)})^2} = - \frac{1}{z} \frac{d}{dt} \frac{1}{1+ z \int_0^t ds e^{2 X(s)}}. \end{equation}
The second is a consequence of the first. The third is obtained from the first, also consistent with the fact that the inverse path transform of $T_z$ is $T_{-z}$.

\item[(ii)] The non-trivial identity in law, from \cite{dufresne2001affine,matsumoto2001relationship},  valid for $\mu<0$, 
\be  \label{idelaw} 
\left( \{ {\sf B}_\mu \}_{t \geq 0} \, ,\, \frac{1}{\int_0^{+\infty} ds e^{2 {\sf B}_\mu(s) } } \right) = 
\left( \{ T_{2 \gamma_{-\mu}}({\sf B}_{-\mu}) \}_{t \geq 0}\, , \, 2 \gamma_{- \mu}  \right)
\ee 
%Using this one shows that
%\be 
%\mathbb{E}[ F[ {\sf B}_s^{-\nu} , s \leq t] \mathbb{E}_{\gamma_b}[ 
%\frac{1}{(2 \gamma_b \int_0^t ds e^{2{\sf B}^{-\nu}(s)} + e^{2 {\sf B}^{-\nu}(t)})^{b-\nu} } ] = 
%\mathbb{E}[ F[T_{2 \gamma_b} ({\sf B}_{\nu}(s)) , s \leq t] ]
%\ee 
\end{enumerate} 
Using these two inputs, the derivation of \eqref{Y} goes as follows. The simple change of drift 
\be
e^{- \frac{1}{2} \int_0^t \left(\frac{dX(s)}{ds}- \nu\right)^2 }= e^{- \frac{1}{2} (\nu^2 - b^2) t } e^{-\frac{1}{2} \int_0^t \left(\frac{dX(s)}{ds}- b\right)^2  + (\nu-b) X(t) } 
\ee 
gives for any observable
\be
\mathbb{E}\left[ O({\sf B}_\nu) \right] = e^{- \frac{1}{2} (\nu^2 - b^2) t } \mathbb{E}\left[ e^{(\nu-b) {\sf B}_b(t) }  O({\sf B}_b) \right].
\ee 
Let us apply it with the choice $O(X):=F[T_{2 \gamma_b}(X)]$. One obtains
\begin{align} 
 \mathbb{E}\left[ F[T_{2 \gamma_b}({\sf B}_\nu)] \right] &= e^{- \frac{1}{2} (\nu^2 - b^2) t } \mathbb{E}\left[ e^{(\nu-b) {\sf B}_b(t)}  F[T_{2 \gamma_b}({\sf B}_b)] \right] \\
& = e^{- \frac{1}{2} (\nu^2 - b^2) t } \mathbb{E}\left[ \frac{e^{(\nu-b) T_{2 \gamma_b}({\sf B}_b)(t)}}{
(1- 2 \gamma_b \int_0^t ds e^{2 T_{2 \gamma_b}({\sf B}_b)(s)}  )^{\nu-b} }
F[T_{2 \gamma_b}({\sf B}_b)] \right] \\
& = e^{- \frac{1}{2} (\nu^2 - b^2) t } \mathbb{E}\left[ e^{(\nu-b) {\sf B}_{-b}(t)}
\left(1-\frac{\int_0^{t} ds e^{2 {\sf B}_{-b}(s)} }{\int_0^{+\infty} ds e^{2 {\sf B}_{-b}(s)} }  \right)^{b-\nu} 
F[{\sf B}_{-b})] \right] \label{last} 
\end{align} 
In the second line we have used the last identity in \eqref{identities}. In the third line the identity in law \eqref{idelaw} has been used
to replace the joint average over $T_{2 \gamma_b}({\sf B}_b)$ and $\gamma_b$ with the same joint average over ${\sf B}_{-b}$
and $1/\int_0^{+\infty} ds e^{2 {\sf B}_{-b}}(s)$. Now since we are interested only in the first part of the process (up to time $t$)
we can replace $\int_t^{+\infty} ds e^{2 {\sf B}_{-b}(s) } \overset{(d)}{=}e^{2 {\sf B}_{-b}(t)}/(2 \gamma_{b})$ thus
\be 
1-\frac{\int_0^{t} ds e^{2 {\sf B}_{-b}(s)} }{\int_0^{+\infty} ds e^{2 {\sf B}_{-b}(s)} }
= \frac{\int_t^{+\infty} ds e^{2 {\sf B}_{-b}(s)} }{\int_0^{t} ds e^{2 {\sf B}_{-b}(s) } + \int_t^{+\infty} ds e^{2 {\sf B}_{-b}(s) } }
\overset{(d)}{=} \frac{e^{2 {\sf B}_{-b}(t)} }{ 2 \gamma_{b} \int_0^{t} ds e^{2 {\sf B}_{-b}(s) } + e^{2 {\sf B}_{-b}(t)} } 
\ee 
Inserting into \eqref{last} and using again the CM theorem to relate expectations over ${\sf B}_{-b}$ to ${\sf B}_{-\nu}$ one 
finally obtains the identity
\begin{equation} 
\mathbb{E}[ F[T_{2 \gamma_b}({\sf B}_\nu) , s \leq t] ]
%&=& e^{- \frac{1}{2} (\nu^2 - b^2) t } \mathbb{E}[ e^{(b-\nu) {\sf B}_{-b}(t)} 
%(\frac{1 }{ 2 \gamma_{-b} \int_0^{t} ds e^{2 {\sf B}_{-b}(s) } + e^{2 {\sf B}_{-b}(t)} })^{b-\nu} F[{\sf B}_{-b}) , s \leq t ] ] \nonumber \\
= \mathbb{E}\left[  \left(\frac{1 }{ 2 \gamma_{b} \int_0^{t} ds e^{2 {\sf B}_{-\nu}(s) } + e^{2 {\sf B}_{-\nu}(t)} }\right)^{b-\nu} F[{\sf B}_{-\nu} , s \leq t] \right] 
\end{equation}  
which for $\nu=0$ and $b=u$ and upon the correspondence \eqref{correspondence1},\eqref{correspondence2} leads to \eqref{Y}.

%Note the identity in law
%\be 
%e^{-2 {\sf B}_\mu(t)} \int_0^t ds e^{2 {\sf B}_\mu(s)} = \int_0^t ds e^{2 {\sf B}_{-\mu}(s)}
%\ee 

\bibliographystyle{eplbib_withtitles}
\bibliography{biblio1.bib}